\documentclass[]{pasj01}
\usepackage{lineno}
\usepackage{url}
\usepackage{threeparttable}
\usepackage{array}
\usepackage{xcolor}

\newcommand{\nar}{NewAR} 

\newcommand{\ulumi}{erg s$^{-1}$}
\newcommand{\uflux}{erg cm$^{-2}$ s$^{-1}$}
\newcommand{\ucolumn}{cm$^{-2}$}

\begin{document} 
\Received{}
\Accepted{}

\title{
Discovery of a new supergiant fast X-ray transient MAXI J0709$-$159
associated with the Be star LY CMa
}


\author{
Mutsumi \textsc{Sugizaki}\altaffilmark{1}}

\author{
Tatehiro \textsc{Mihara}\altaffilmark{2}}

\author{
Kohei \textsc{Kobayashi}\altaffilmark{3}}

\author{
Hitoshi \textsc{Negoro}\altaffilmark{3}}

\author{
Megumi \textsc{Shidatsu}\altaffilmark{4}}

\author{
Sean N. \textsc{Pike}\altaffilmark{5}}

\author{
Wataru \textsc{Iwakiri}\altaffilmark{6}}

\author{
Sota \textsc{Urabe}\altaffilmark{6}}

\author{
Motoko \textsc{Serino}\altaffilmark{7}}

\author{
Nobuyuki \textsc{Kawai}\altaffilmark{8}}

\author{
Motoki \textsc{Nakajima}\altaffilmark{9}}

\author{
Jamie A. \textsc{Kennea}\altaffilmark{10}}

\author{
Zhu \textsc{Liu}\altaffilmark{11}}

\altaffiltext{1}{National Astronomical Observatories, Chinese Academy of Sciences, 20A Datun Rd, Beijing 100012, China}

\altaffiltext{2}{RIKEN, 2-1 Hirosawa, Wako, Saitama 351-0198, Japan}

\altaffiltext{3}{Department of Physics, Nihon University, 1-8 Kanda Surugadai, Chiyoda-ku, Tokyo, 101-8308, Japan} 

\altaffiltext{4}{Department of Physics, Ehime University,  2-5, Bunkyocho, Matsuyama, Ehime 790-8577, Japan}

\altaffiltext{5}{Cahill Center for Astronomy and Astrophysics, California Institute of Technology, Pasadena, CA 91125, USA}

\altaffiltext{6}{Department of Physics, Chuo University, 1-13-27 Kasuga, Bunkyo-ku, Tokyo 112-8551, Japan}

\altaffiltext{7}{Department of Physical Sciences, Aoyama Gakuin University, 5-10-1 Fuchinobe, Chuo-ku,  Sagamihara, Kanagawa, 252-5258, Japan} 

\altaffiltext{8}{Department of Physics, Tokyo Institute of Technology,  2-12-1 Ookayama, Meguro-ku, Tokyo 152-8551, Japan}

\altaffiltext{9}{School of Dentistry at Matsudo, Nihon University,  2-870-1 Sakaecho-nishi, Matsudo, Chiba 101-8308, Japan}

\altaffiltext{10}{Department of Astronomy and Astrophysics, Pennsylvania State University, 525 Davey Laboratory, University Park, PA 16802, USA}

\altaffiltext{11}{Max Planck Institute for Extraterrestrial Physics, Giessenbachstrasse 1, 85748, Garching, Germany}

\email{mutsumi@nao.cas.cn, tmihara@riken.jp}


\maketitle

\begin{abstract}
  We report on the discovery of a new supergiant fast X-ray transient (SFXT),
  MAXI J0709$-$159, and its identification with LY CMa
  (also known as HD 54786).  On 2022 January 25, a new flaring X-ray
  object named MAXI J0709$-$159, was detected by Monitor of All-sky
  X-ray Image (MAXI).
  Two flaring activities were observed in the two scans of $\sim 3$ hours apart, 
  where the 2--10
  keV flux reached $5\times 10^{-9}$ erg cm$^{-2}$ s$^{-1}$.  During
  the period, the source exhibited a large spectral change suggesting
  that the absorption column density $N_\mathrm{H}$ increased from
  $10^{22}$ cm$^{-2}$ to $10^{23}$ cm$^{-2}$.
  NuSTAR follow-up observation on January 29 identified a new X-ray
  source with a flux of $6\times 10^{-13}$ erg cm$^{-2}$ s$^{-1}$ 
  at the position consistent with LY CMa, which has been identified as
  B supergiant as well as Be star, 
  located at the 3 kpc distance.
  The observed X-ray activity characterized by the short ($\lesssim$ several
  hours) duration, the rapid ($\lesssim$ a few seconds) variabilities
  accompanied with spectral changes, and the large luminosity swing
  ($10^{32}$--$10^{37}$ erg s$^{-1}$) agree with those of SFXT.
  On the other hand, optical spectroscopic observations of LY CMa
  revealed a broad $H\alpha$ emission line,
  which may indicate the existence of a Be circumstellar disk.
  These obtained results suggest that the optical companion, LY CMa, certainly 
  has a complex circumstellar medium including dense clumps.
\end{abstract}


\section{Introduction} 

MAXI (Monitor of All-sky X-ray Image: \cite{Matsuoka2009}) 
on the International Space Station (ISS)
has been continuously scanning the almost entire sky  
every ISS orbital cycle ($\sim$92 minutes)
since the in-orbit operation started in 2009.
It provides us a unique opportunity 
to discover new objects and study their transient behaviors. 
In fact, we have discovered 31 new X-ray objects
including 14 black-hole binaries, 13 neutron-star binaries,
and one white-dwarf binary,
that appeared in our Galaxy and the Small Magellanic Cloud (e.g. \cite{Mihara2022Handbook}).
The data also enable us to study their variabilities 
in time scales from hours to over the 12 years. 
To uncover the nature of these new transient objects, 
prompt follow-up observations with large-area X-ray telescopes 
as well as multi-wavelength observations from both space and ground observatories 
are essential.
The MAXI nova-alert system \citep{Negoro2016PASJ}
and prompt coordinated observations with Swift, NICER, NuSTAR satellites, 
have been working effectively 
for these transient studies 
such as an ignition of a classical nova, MAXI J0158$-$744 \citep{2013ApJ...779..118M}, 
a new Be X-ray binary pulsar, MAXI J0903$-$531 \citep{2021arXiv210806365T}, and
a faint and short-duration back-hole binary candidate,
MAXI J1848$-$015 \citep{2022ApJ...927..190P}.

On 2022 January 25, 
MAXI discovered a new X-ray 
transient with an instantaneous 4--10 keV flux of 270 mCrab
($\sim 5\times 10^{-9}$ {\uflux}), 
named MAXI J0709$-$159 (hereafter MAXI J0709), 
in the constellation Canis Majoris \citep{ATel15178.S}. 
The source was first detected during the scan transit at UT 10:42,
but not detected in the next scan transit at UT 12:15. 
However, it was detected again in the other next scan at UT 13:48 \citep{ATel15188.K}.
This means that the source exhibited a large intensity variation
within the 3 hours.
%
%
Three hours after the MAXI discovery, NICER (The Neutron star Interior
Composition ExploreR; \cite{Gendreau2012}) started multiple pointing
observations covering almost the entire region of the MAXI position
error (circle of radius $\sim 0\fdg 2$).
The new transient was successfully detected with the pointing observation 
carried out 6 minutes after the MAXI scan at UT 13:46.
The NICER observations refined the source position with the error of $3\arcmin$
and also revealed that the X-ray flux had declined by a factor of $\sim10$ from the MAXI observations \citep{ATel15181.I}.
%
%

From 2022 January 19 to February 18, the Swift X-ray satellite \citep{2004ApJ...611.1005G} stopped the normal operation
because the attitude control system was troubled\footnote{\url{https://swift.gsfc.nasa.gov/news/2022/safe_mode.html}}.
Hence, we applied for the NuSTAR (Nuclear Spectroscopic Telescope Array; \cite{Harrison2013}) 
ToO (Time of Opportunity) observation.
The NuSTAR observation was carried out 
on 2022 January 29, 4 days after the discovery.
The result revealed a new point source within the error circle of the MAXI J0709 position uncertainty.
The position of the new source is consistent with LY CMa 
(also known as HD 54786),
which has been identified as a Be star \citep{ATel15193.N,ATel15194.N}.
In Gaia Early Data Release 3,
the distance is estimated to be 
$D=3.03^{+0.31}_{-0.27}$ kpc 
\citep{2021AJ....161..147B}\footnote{
\url{https://vizier.u-strasbg.fr/viz-bin/VizieR-3?-source=I/352}
}.
%
From optical follow-up observations and archival data analysis,
\citet{2022arXiv220604473B} suggested that
LY CMa would be an evolved Be star, rather than a main sequence star or a supergiant.
%
A radio observation was carried out on 2022 January 31 with the MerrKAT radio telescope,
but no significant emission was detected at 1.28 GHz
and the 3-$\sigma$ upper limit was estimated to be 57 $\mu$Jy
\citep{2022ATel15209....1R} 
%
Finally, the Swift ToO observation was carried out on 2022 February 23.

In this paper, we report the discovery of the new transient MAXI J0709
and results of the MAXI, NuSTAR, Swift, and eROSITA observations
of the identified X-ray object.
We also report an optical follow-up observation of the optical counterpart
in the Chuo University.
Based on the obtained results, we discuss the nature of the new X-ray object.
In the following throughout the paper, errors represent 90\%
confidence limits of statistical uncertainties unless otherwise
specified.

\section{MAXI observations and data analysis}

MAXI Gas Slit Camera (GSC; \cite{Mihara2011,Sugizaki2011})
consists of 12 identical camera units, 
namely GSC\_0, ..., GSC\_9, GSC\_A, GSC\_B.
Utilizing the two wide ($160\arcdeg\times3\arcdeg$) FOVs,
GSC scans the whole sky
every 92 minutes.
We investigated X-ray activities of a new transient MAXI J0709
in time scales shorter than each scan transit ($\sim 40$ s) and
longer than the scan cycle ($\sim 92$ minutes),
using the GSC data.
We here employed GSC event files of the process version 2.1, taken via the low-speed telemetry interface,
and performed the data analysis using MAXI software included in HEASoft version 6.29 
the calibration database (CALDB) version 20210504 released from JAXA data
archive\footnote{https://www.darts.isas.jaxa.jp/astro/maxi/data.html}.

\subsection{GSC light curve}

\begin{figure*}
  \begin{center}
  \includegraphics[width=0.4\textwidth]{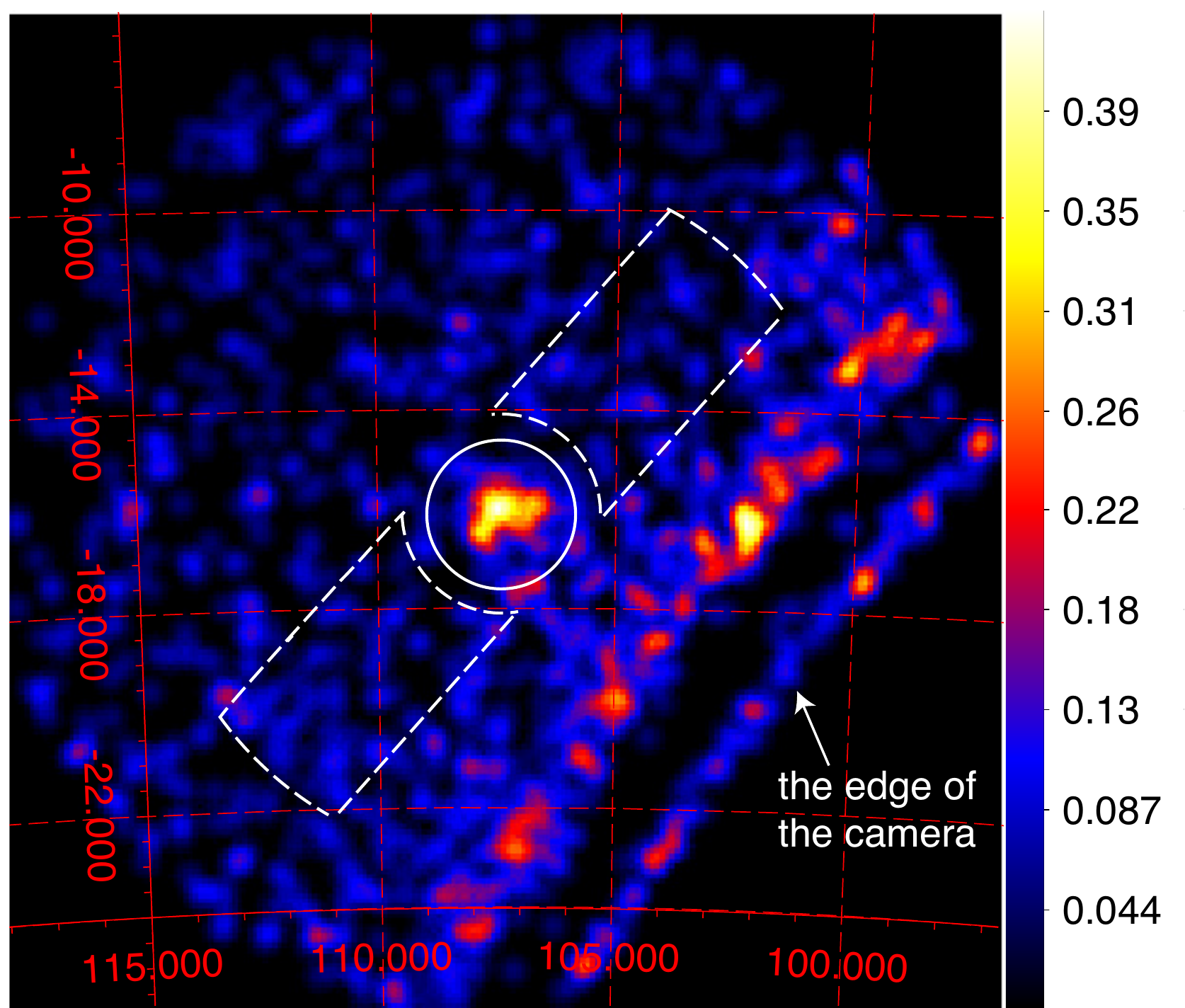}
  \hspace{5mm}
  \includegraphics[width=0.4\textwidth]{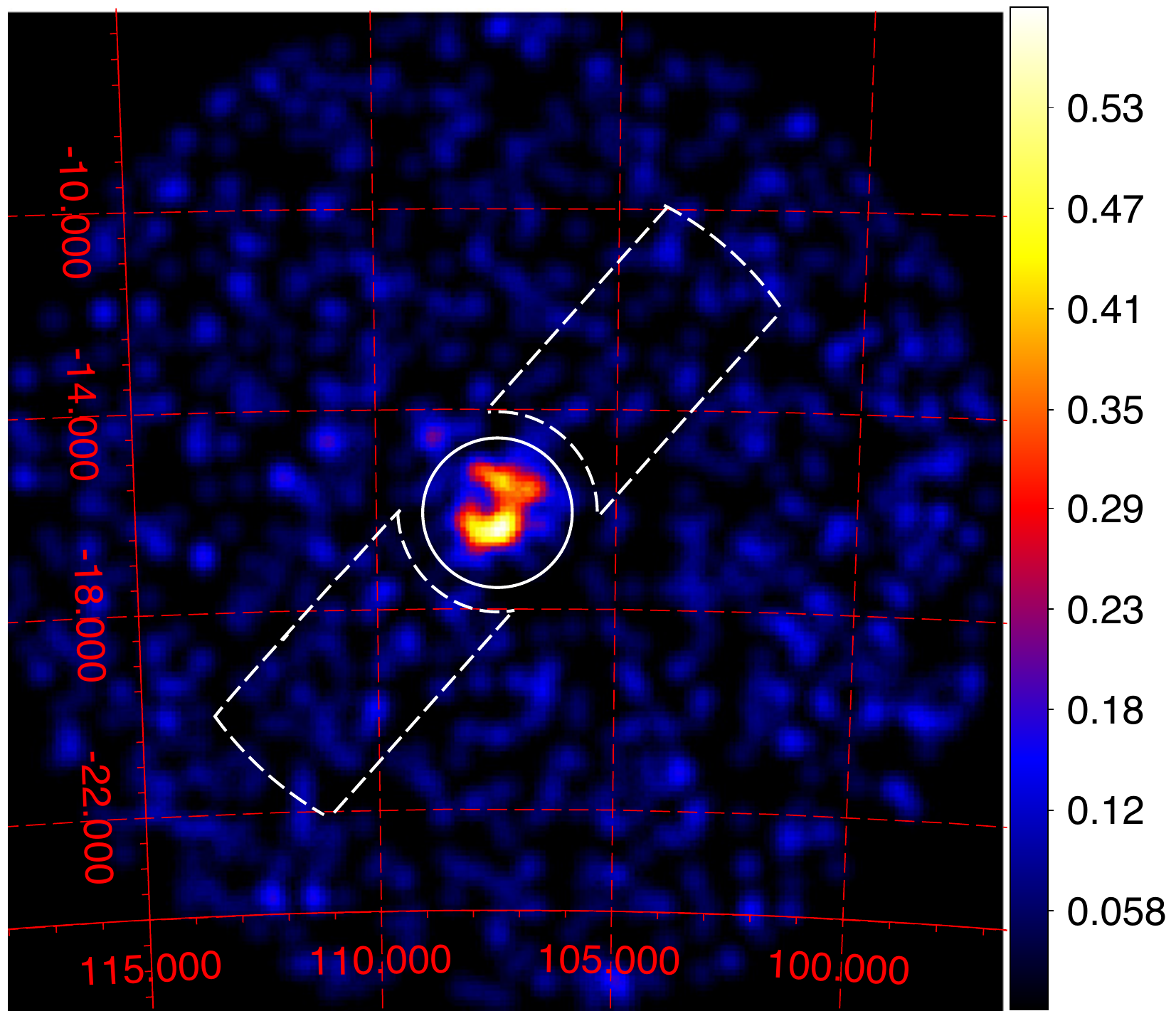}
  \end{center}
  \caption{
    GSC 2--20 keV images obtained by GSC\_4 (left) and GSC\_5 (right) units,
    within $10\arcdeg$ of MAXI J0709
    at the Scan-A (UT 10:39-10:46 on January 25).
    The image is smoothed with a Gaussian kernel of $\sigma =2\arcdeg$.
    The source and background regions used in spectral analysis
    are shown by the solid and dashed lines, respectively.
  }
\label{fig:koba_gsc_image}
\end{figure*}

We extracted GSC light curves of MAXI J0709  by fitting each GSC scan image
with a model consisting of 
a point-spread function (PSF) for the target source and a uniform background
\citep{Morii2016}.
The source position was fixed at
the position parameters refined by the NuSTAR observation
(section \ref{sec:nustar_image}).
Figure \ref{fig:koba_gsc_image} shows
GSC 2--20 keV images taken by the GSC\_4 and GSC\_5 units,
when MAXI J0709 was first detected
on 2022 January 25 (MJD 59604).
Within 10 days before and after the first detection, 
the source had been observed
by either or both of these two GSC units.

\begin{figure*}
  \begin{center}
  \includegraphics[width=0.7\textwidth]{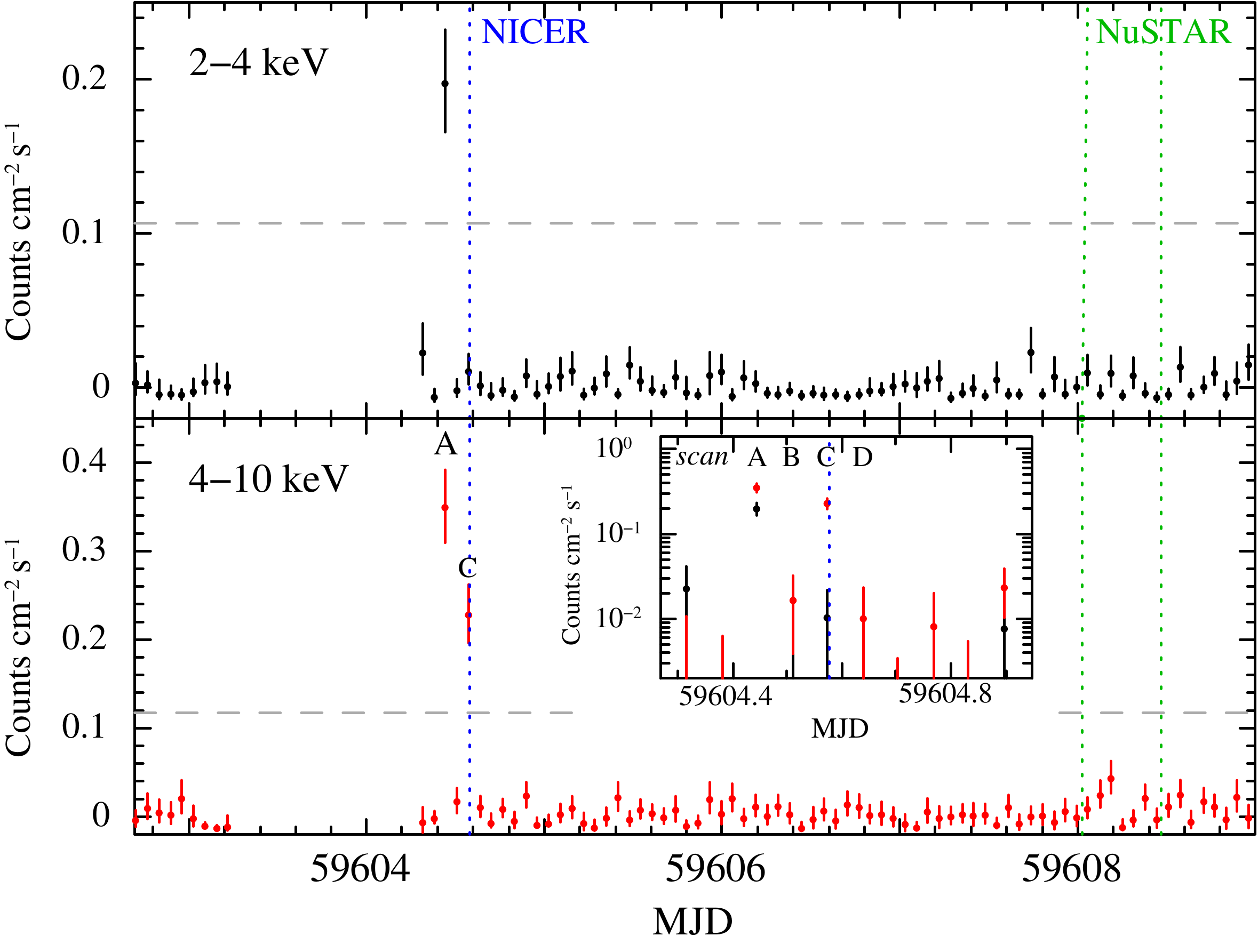}
  \end{center}
  \caption{
  GSC light curves of MAXI J0709 
  in the 2--4 keV (black) and 4--10 keV (red) bands.
  Horizontal dashed lines represent the expected count rates for a 100 mCrab source, 
  0.106 counts cm$^{-2}$ s$^{-1}$ in 2--4 keV and 
  0.117 counts cm$^{-2}$ s$^{-1}$ in 4--10 keV.
  Vertical dotted lines represent the epoch of the NICER observation start (blue)
  and the NuSTAR observation period (green).
  The inset is a log-linear plot of the 0.7 d around the source onset, where
  Scan-A, B, C, D are annotated.
  }
  \label{fig:maxi_lc}
  \end{figure*}

Figure \ref{fig:maxi_lc} shows the obtained 
2--4 keV and 4--10 keV light curves
from the 2 days before to the 5 days after the first detection.
The data gap from MJD 59603.2 to 59604.3
corresponds to the period when the source position
was shadowed by the frame structure 
in the GSC detector\footnote{
Also see the light curve in the MAXI homepage
\url{http://maxi.riken.jp/pubdata/v7lrkn/J0709-161/}.
}.
After the data gap, there was no significant flux in the two scans.
Then, the first X-ray activity was detected in the scan transit
at UT 10:42 (MJD 59604.446, Scan-A). 
The source count rates averaged over the scan transit were 
$0.19^{+0.03}_{-0.03}$ counts\,cm$^{-2}$\,s$^{-1}$ (180 mCrab) in 2--4 keV and 
$0.35^{+0.04}_{-0.04}$  counts\,cm$^{-2}$\,s$^{-1}$ (300 mCrab) in 4--10 keV. 
In the next scan transit at UT 12:15 (MJD 59604.510, Scan-B), 
there was no significant flux over the background with
the $1\sigma$ upper limits of
0.005 counts\,cm$^{-2}$\,s$^{-1}$ (5 mCrab) in 2--4 keV 
and 0.032 counts\,cm$^{-2}$\,s$^{-1}$ (27 mCrab) in 4--10 keV.
In the other next scan transit at UT 13:48 (MJD 59604.575, Scan-C), 
the source was detected again but only in the 4--10 keV band
with $0.23^{+0.03}_{-0.03}$  counts\,cm$^{-2}$\,s$^{-1}$ (200 mCrab).
Those mean that the source intensity changed every scan transit 
by a factor of $\gtrsim 10$ and also the spectrum changed.
In the following scans after MJD 59604.639 (Scan-D), the source activity
went down below the GSC sensitivity limit.

\begin{figure}
\begin{center}
\includegraphics[width=0.48\textwidth]{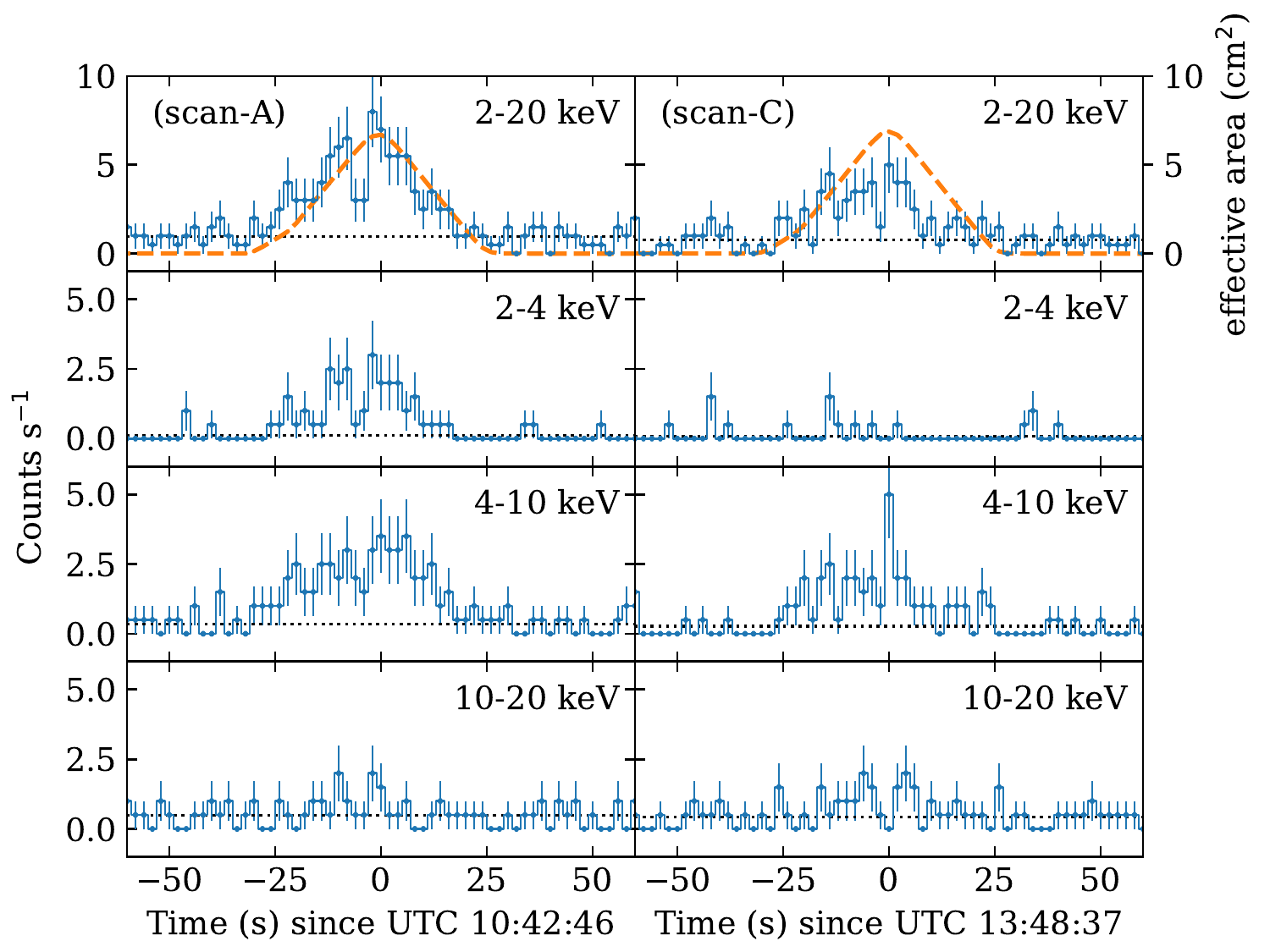} 
\end{center}
\caption{
GSC count rates during the scan transits centered at UT 10:42 (Scan-A) and 13:48 (Scan-C)
on 2022 January 25.
Data are binned every 2 s. 
Dotted lines represent the background levels.
Dashed lines in the top panels show the effective-area variation
for the MAXI J0709 sky position. 
}
\label{fig:scan_lc}
\end{figure}

We then investigated the source variability 
within each scan transit of Scan-A and Scan-C, which include significant source photons.
Figure \ref{fig:scan_lc} shows count rate variations 
during Scan-A and Scan-C,
compared with the effective-area variations for the source position
and the backgrounds estimated from 
the data in the adjacent source-free region.
In Scan-A, the 2--20 keV count rate roughly 
traces the effective-area variation, 
indicating that the photon flux was approximately 
constant during the scan  transit.
However, at the middle of the transit, a dip-like structure
lasting for $\sim 4$ s is clearly seen.
We fitted the data around the dip 
with a constant-flux model, 
i.e. a normalized effective-area curve plus background.
The result gives Cash statistic (C-stat; \cite{Cash1979}) $=7.5$ for 3 degrees of freedom (d.o.f),
meaning that the constant model is rejected with a confidence of $>90$\%.
In Scan-C, the 4--10 keV count rate shows a sharp peak at the center of the scan.
We performed the model fit same as in Scan-A
and obtained the similar result that 
the constant-flux model is rejected with a confidence of $>95$\%. 
These results suggest the source has 
a rapid time variability on time scales of a few seconds or less.

We also investigated the past source activity 
from the beginning of the MAXI in-orbit operation in 2009 August.
At the distance of $0\fdg55$ from the refined MAXI J0709 position,
another X-ray source, 3MAXI J0708-155, was reported
in the 7-year MAXI/GSC source catalog \citep{2018ApJS..235....7H}.
There, the object is identified as the Blazar, PKS 0706-15.
Although this nearby X-ray object can be distinguished from 
MAXI J0709 with the MAXI/GSC position accuracy ($\lesssim 0.2\arcdeg$), 
the PSFs of these two objects have an overlap,
which causes the source confusion.
We carefully estimate the amount of the possible confusion
and confirmed that MAXI J0709 had not shown 
any significance ($> 4\sigma$) flaring activity
until the present event.

\subsection{X-ray spectra of two flaring events}
\label{sec:maxispec}

Next, we analyzed GSC energy spectra in Scan-A and Scan-C in detail.
Source spectra were extracted from a circular region within 
$1\fdg 5$ from the source position.
To avoid high background area near the edge of the detector in GSC\_4 (figure \ref{fig:koba_gsc_image}), 
background spectra were extracted from 
the overlap region between an annulus of inner and outer radii of $2\arcdeg$ and $7\arcdeg$,  
and a $14\arcdeg\times3\arcdeg$ rectangle along the scan direction,
both centered at the source position.
These source and background regions are illustrated
in figure \ref{fig:koba_gsc_image}.
We confirmed that background subtracted spectra 
were consistent between GSC\_4 and GSC\_5, and then added them into GSC\_4$+$GSC\_5 spectra.

\begin{figure*}
\begin{center}
\includegraphics[width=0.45\textwidth]{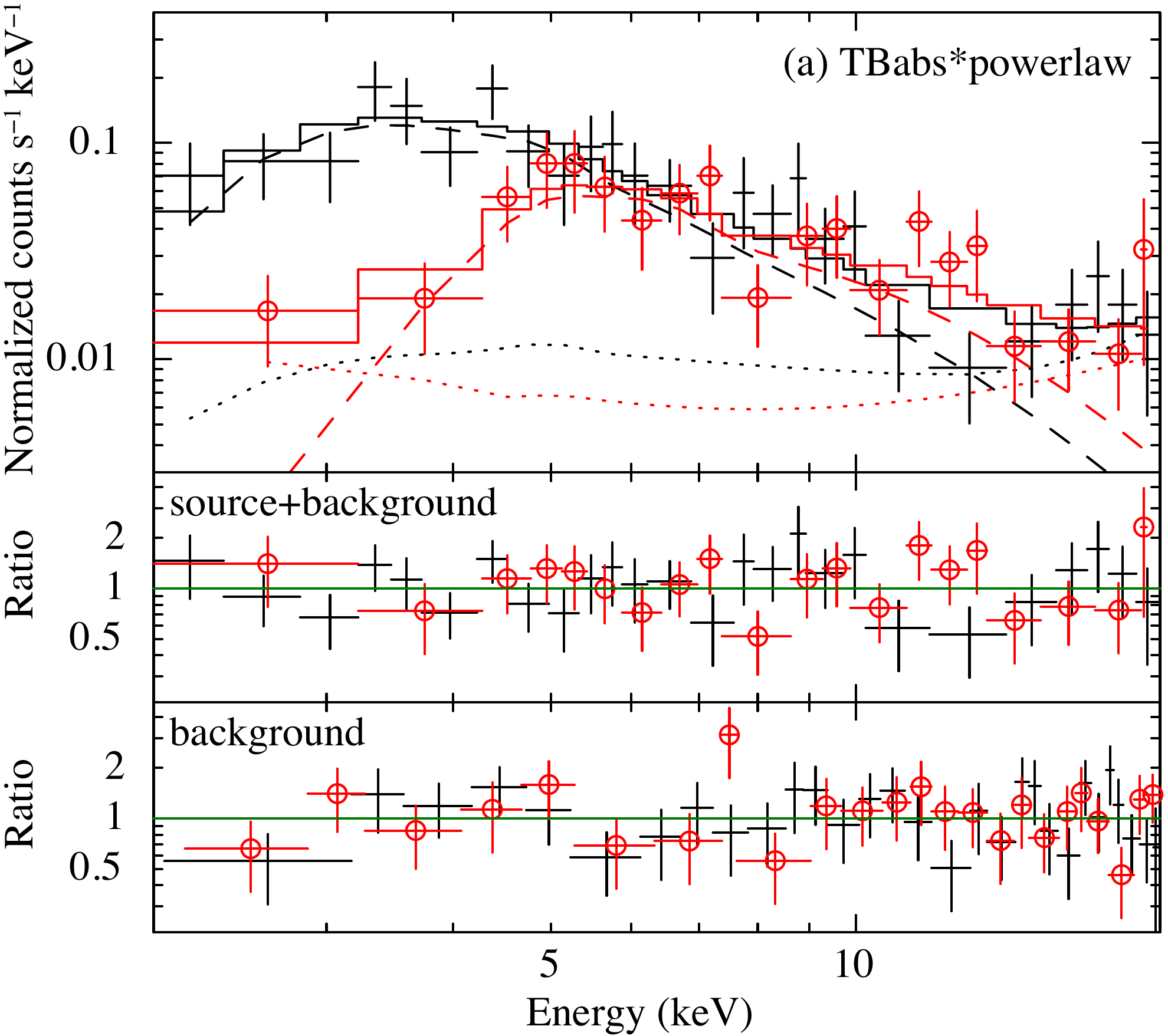}
\hspace{5mm}
\includegraphics[width=0.45\textwidth]{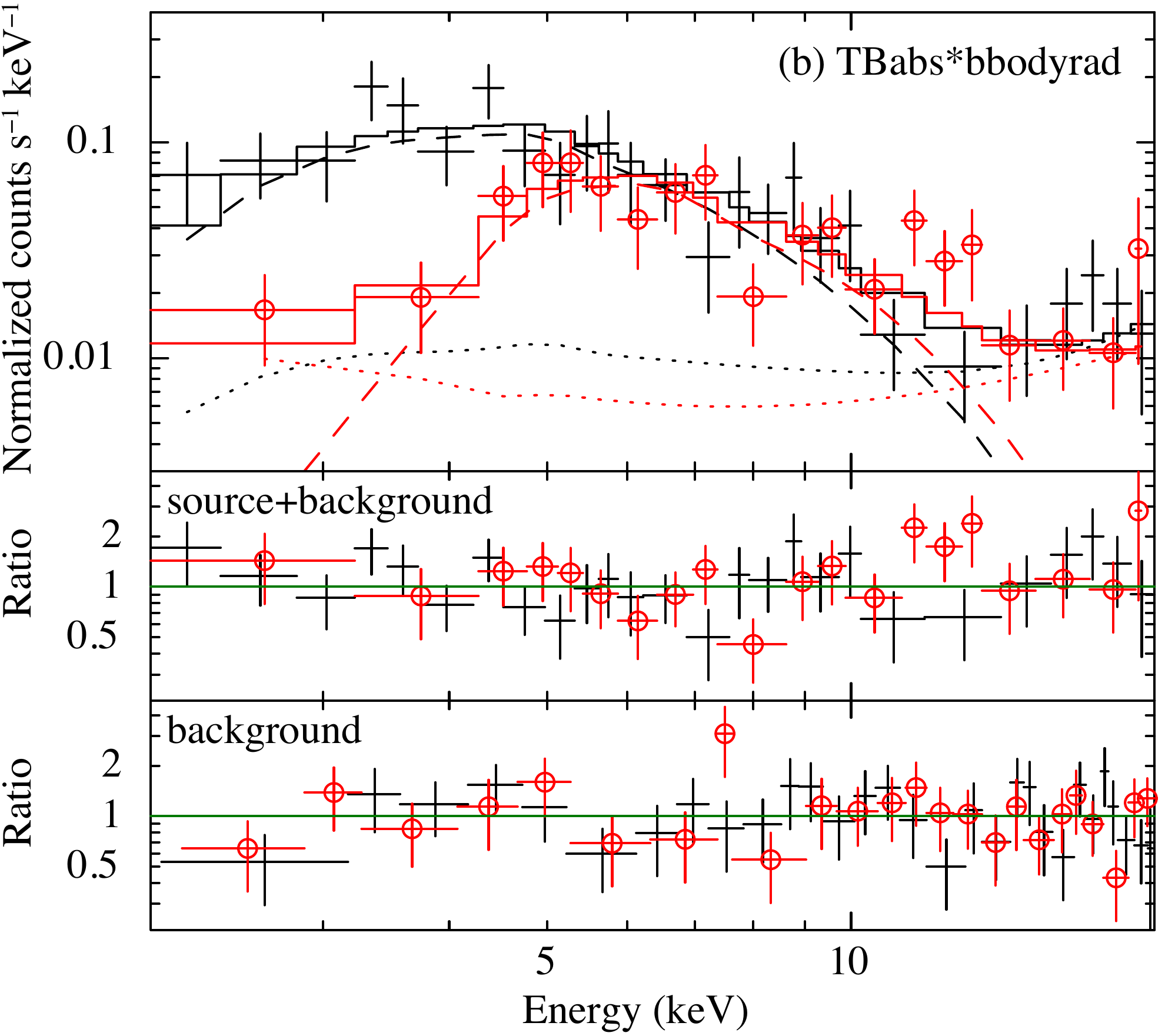}
\end{center}
\caption{
GSC energy spectra 
fitted with (a) powerlaw
and (b) blackbody with an interstellar absorption.
In all panels,
black dots and red circle marks
represent Scan-A and Scan-C data, respectively.
(Top panel) Observed source-region spectra, folded with GSC response function.
Dashed lines, dotted lines, and solid histograms represent
model spectra for source, background, and source$+$background,
respectively.
(Middle panel) Data-to-model ratio of source-region (including background) spectra.
(Bottom panel) Data-to-model ratio of background spectra.
For visual clarity, plotted data are rebinned. 
}
\label{fig:koba_gsc_spec}
\end{figure*}

Figure \ref{fig:koba_gsc_spec} shows the obtained source-region spectra in Scan-A and Scan-C.
As expected from the light curves in figure \ref{fig:maxi_lc},
the Scan-C spectrum has a lower-energy cutoff below 4 keV.
We performed model fitting on XSPEC version 12.11.1
\citep{Arnaud1996} employing the C-statistic method.
The energy response matrix for each data was calculated with {\tt mxgrmfgen} included in HEASoft.
Also, spectral data were binned so that each bin contains at least one event.
We first checked that background spectra were well represented by a two-powerlaw model.
Then, we fitted source-region spectra including backgrounds and background spectra
simultaneously to determine the source spectral model.

Firstly, we fitted each of Scan-A and Scan-C spectra with a power-law with an interstellar-medium (ISM)
absorption.
We hereafter employed the Tuebingen-Boulder ISM absorption model ({\tt TBabs}) with the solar abundances provided by \citet{Wilms2000}.
Both fits to the Scan-A and Scan-C spectra were acceptable.
Table \ref{table:gsc_spectra_par} summarizes the best-fit model parameters.
While the power-law photon indices,
$\Gamma = 2.3^{+0.7}_{-0.6}$ in Scan-A and $2.7^{+1.2}_{-1.0}$ in Scan-C,
are consistent within the errors, 
the absorption column densities,
$N_\mathrm{H} = 5.5^{+5.8}_{-4.8}\times 10^{22}$ {\ucolumn} in Scan-A
and $58^{+53}_{-30}\times 10^{22}$ {\ucolumn} in Scan-C,
are significantly different.
The $N_\mathrm{H}$ value in Scan-A is larger than
the Galactic $\mathrm{H}_\mathrm{I}$ density, $0.5\times 10^{22}$ {\ucolumn},
in the source direction
\citep{2016A&A...594A.116H}.
%
We then fitted the two spectra with a common $\Gamma$ simultaneously.
The fit was accepted similarly, and 
the best-fit $\Gamma=2.4^{+0.6}_{-0.5}$ was obtained. 
In this case, 
the absorption-corrected 2--10 keV fluxes 
are consistent between Scan-A and Scan-C
at $\sim 0.5\times 10^{-8}$ {\uflux}.

We also fitted the two spectra with a blackbody model
({\tt bbodyrad}) with an interstellar absorption.
The fits were acceptable in both Scan-A and Scan-C, as in the same way as the fits with a powerlaw model.
The obtained model parameters are summarized in table \ref{table:gsc_spectra_par}.
There, the blackbody model normalizations are given
by a source radius $R_\mathrm{BB}$.
$N_\mathrm{H}$ of Scan-A was constrained only by the upper limit
($<2.2\times10^{22}$ {\ucolumn}). 
Although the blackbody temperatures,
$T_\mathrm{BB}=1.5^{+0.2}_{-0.2}$ keV in Scan-A and $2.2^{+0.8}_{-0.5}$ keV in Scan-C,
were slightly different,
we tried a simultaneous fit to the two spectra with a common $T_\mathrm{BB}$.
The fit was still acceptable and the best-fit $T_\mathrm{BB}=1.6^{+0.2}_{-0.2}$ keV was obtained.

\newcolumntype{P}[1]{>{\centering\arraybackslash}p{#1}}

\begin{table}
\begin{threeparttable}
\caption{Best-fit spectral parameters of MAXI J0709}
\label{table:gsc_spectra_par}
\begin{tabular}{p{0.2\textwidth}P{0.1\textwidth}P{0.1\textwidth}}
\multicolumn{3}{l}{Model: TBabs*powerlaw} \\
\hline
\ \ Parameter & Scan-A & Scan-C \\
\hline
\ \ $N_\mathrm{H}$ ($10^{22}$ cm$^{-2}$) & $5.5^{+5.8}_{-4.8}$ & $58^{+53}_{-30}$\\
\ \ $\Gamma$ & $2.3^{+0.7}_{-0.6}$ & $2.7^{+1.2}_{-1.0}$ \\
\ \ $F_{\mathrm{abs}}$$^{a}$ & $0.48^{+0.08}_{-0.07}$ & $0.29^{+0.07}_{-0.06}$\\
\ \ $F_{\mathrm{unabs}}$$^{b}$ & $0.7^{+0.4}_{-0.2}$ & $1.8^{+3.6}_{-1.1}$\\
\ \ C-stat/d.o.f & 202/267 & 177/210 \\
\hline
\ \ $N_\mathrm{H}$ ($10^{22}$ cm$^{-2}$) & $6.2^{+5.3}_{-4.4}$ & $51^{+27}_{-20}$ \\
\ \ $\Gamma$ &  \multicolumn{2}{c}{$2.4^{+0.6}_{-0.5}$} \\
\ \ $F_{\mathrm{abs}}$$^{a}$   &$0.48^{+0.08}_{-0.07}$ & $0.29^{+0.06}_{-0.06}$ \\ 
\ \ $F_{\mathrm{unabs}}$$^{b}$  &$0.7^{+0.3}_{-0.2}$ & $1.4^{+1.4}_{-0.6}$ \\ 
\ \ C-stat/d.o.f &   \multicolumn{2}{c}{379/478} \\
\hline
\\
\multicolumn{3}{l}{Model: TBabs*bbodyrad} \\
\hline
\ \ Parameter & Scan-A & Scan-C \\
\hline
\ \ $N_\mathrm{H}$ ($10^{22}$ cm$^{-2}$) & $< 2.2$ &  $27^{+32}_{-21}$ \\
\ \ $T_{\mathrm{BB}}$ (keV) &$1.5^{+0.2}_{-0.2}$ & $2.2^{+0.8}_{-0.5}$\\
\ \ $R_{\mathrm{BB}}$$^{c}$ (km) & $3.1^{+0.9}_{-0.7}$ & $1.8^{+1.5}_{-0.9}$\\ 
\ \ $F_{\mathrm{abs}}$$^{a}$ & $0.50^{+0.09}_{-0.08}$& $0.31^{+0.07}_{-0.06}$\\ 
\ \ $F_{\mathrm{unabs}}$$^{b}$ & $0.50^{+0.09}_{-0.08}$& $0.6^{+0.5}_{-0.2}$\\ 
\ \ C-stat/d.o.f & 202/267 & 179/210\\ 
\hline
\ \ $N_\mathrm{H}$ ($10^{22}$ cm$^{-2}$) & $< 1.5$ &  $48^{+32}_{-22}$ \\
\ \ $T_{\mathrm{BB}}$ (keV) & \multicolumn{2}{c}{$1.6^{+0.2}_{-0.2}$}\\
\ \ $R_{\mathrm{BB}}$$^{c}$ (km) & $2.8^{+0.7}_{-0.6}$ & $3.7^{+2.1}_{-1.2}$\\ 
\ \ $F_{\mathrm{abs}}$$^{a}$ & $0.51^{+0.08}_{-0.08}$& $0.32^{+0.07}_{-0.06}$\\ 
\ \ $F_{\mathrm{unabs}}$$^{b}$  & $0.51^{+0.08}_{-0.08}$& $0.9^{+0.7}_{-0.4}$\\ 
\ \ C-stat/d.o.f &   \multicolumn{2}{c}{385/478} \\ 
\hline
\end{tabular}
\begin{tablenotes}
\item[a] Absorbed 2--10 keV flux in $10^{-8}$ erg cm$^{-2}$ s$^{-1}$
\item[b] Unabsorbed 2--10 keV flux in $10^{-8}$ erg cm$^{-2}$ s$^{-1}$
\item[c] Blackbody radius in km. Distance of 3 kpc assumed.

\end{tablenotes}
\end{threeparttable}
\end{table}

\section{NuSTAR observation and data analysis} 

The NuSTAR observation of MAXI J0709 was carried out in 2022 January
29 UT 00:21--11:21 (OBSID: 90801304002), with a net exposure of 18 ks.
We performed data analysis using NuSTAR Data Analysis Software {\tt nustardas} version 2.1.1 
included in HEASoft version 6.29 and the CALDB version 20220215, following the NuSTAR Data Analysis Quick Start
Guide\footnote{\url{https://heasarc.gsfc.nasa.gov/docs/nustar/analysis/nustar_quickstart_guide.pdf}}
and the NuSTAR Data Analysis Software
Guide\footnote{\url{https://heasarc.gsfc.nasa.gov/docs/nustar/analysis/nustar_swguide.pdf}}.
The unfiltered event files were first reprocessed by {\tt nupipeline}. 
According to the recommendations given by the NuSTAR team, 
we screened high background-rate data taken 
during the South Atlantic Anomaly (SAA) passages 
by combining options SAAMODE$=$optimized and TENTACLE$=$yes 
with the default SAA calculation option (saacalc$=$3).
The resultant cleaned event files were used to obtain the image, light
curve and time-averaged spectrum.

\subsection{Localization and optical identification with NuSTAR image}
\label{sec:nustar_image}

\begin{figure*}
\begin{center}
\includegraphics[width=0.42\textwidth]{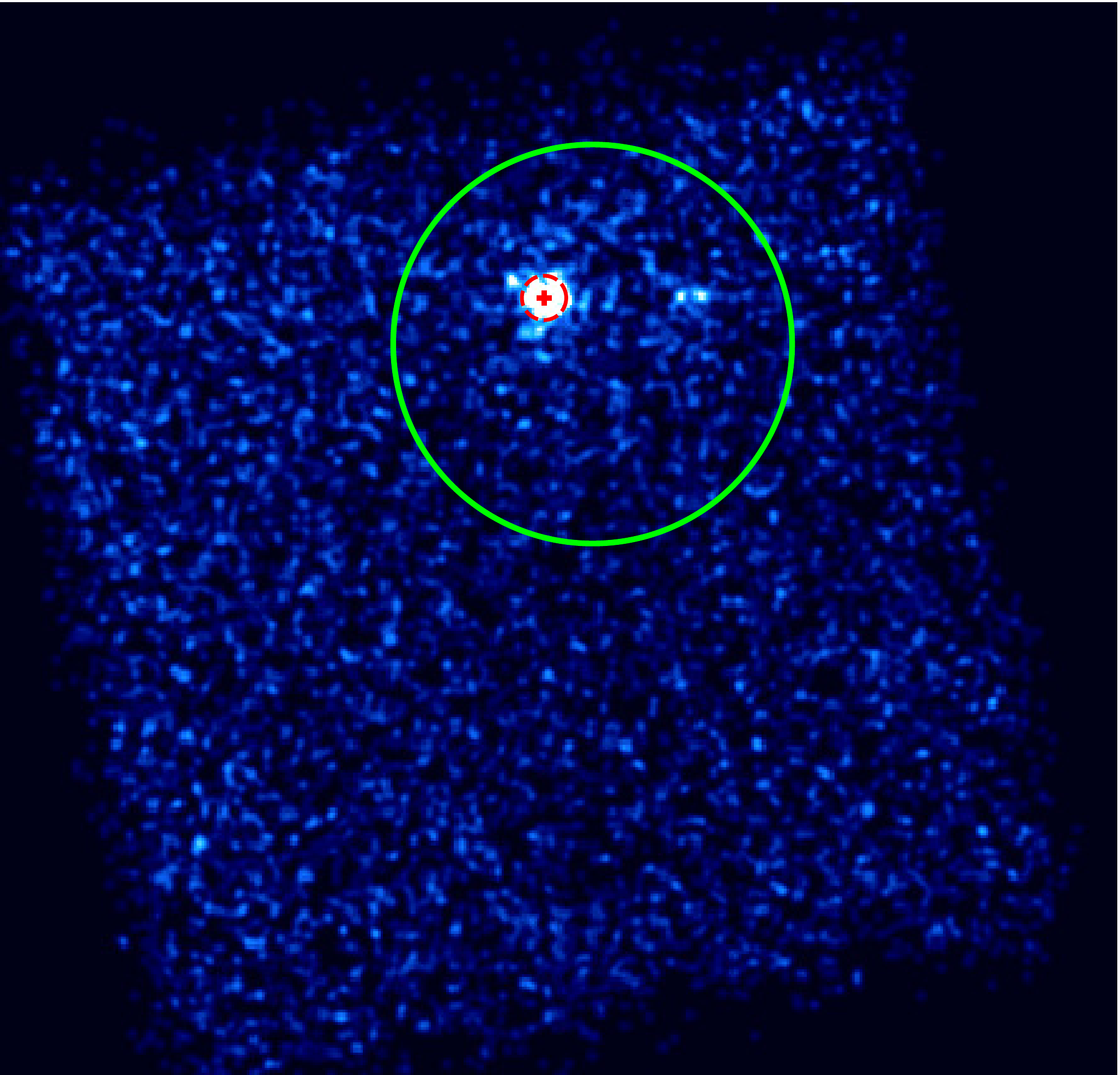}
\includegraphics[width=0.468\textwidth]{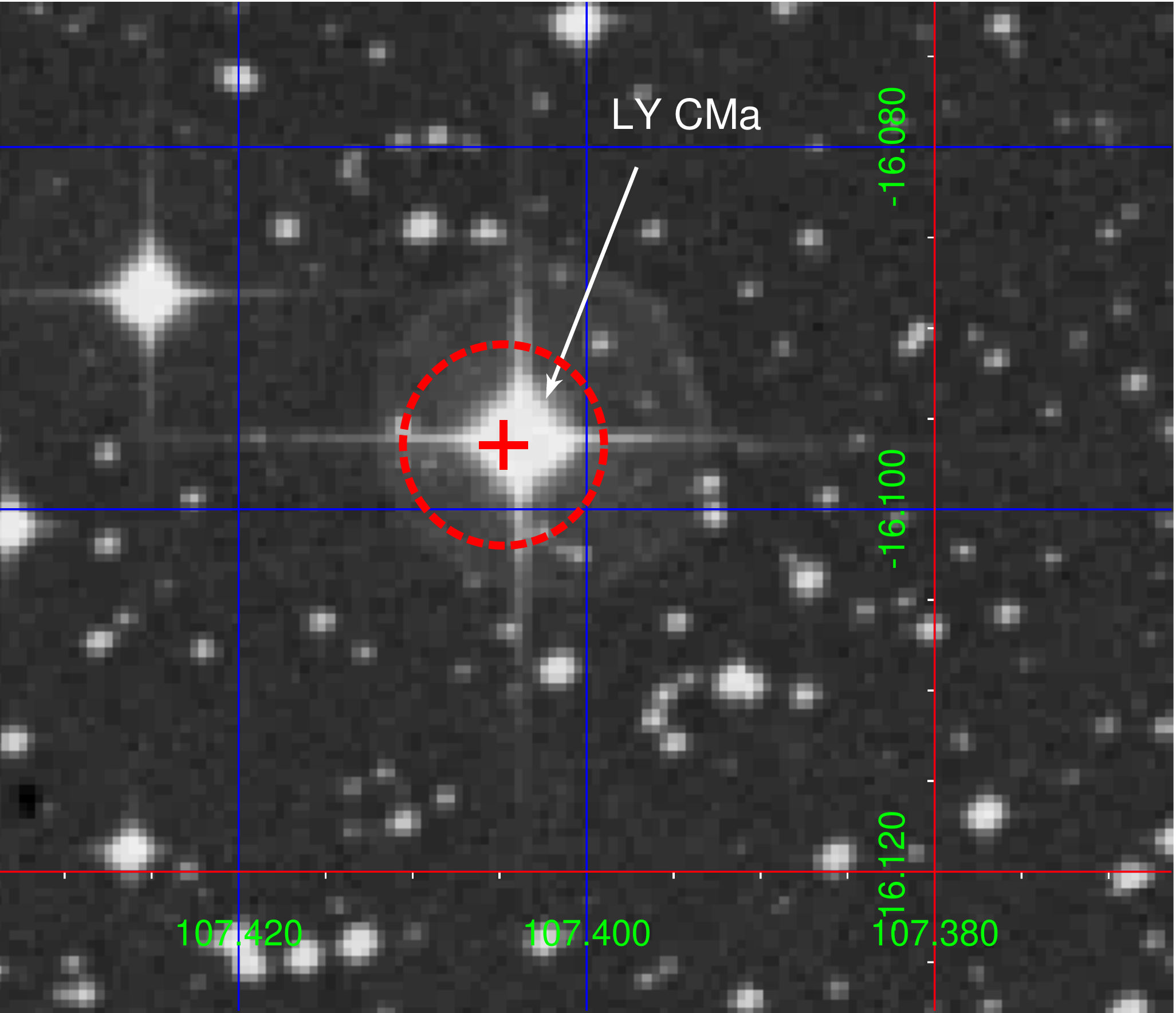}
\end{center}
\caption{
  (Left) NuSTAR/FPMA$+$FPMB 3--20 keV image around the source, where
  Gaussian smoothing is applied to the image with a radius of 2 pixels. 
  The red cross and dashed circle indicate the best-fit position and the
  error (\timeform{20"}, including the systematic error) determined with NuSTAR, 
  respectively.  
  The NICER error circle (with a \timeform{3'} radius) is also shown with the green line. 
  (Right) Optical image of the DSS Original Digitized Sky Survey.
  The NuSTAR source position and error are shown with the red cross 
  and dashed circle. 
}  
\label{fig:nuimg}
\end{figure*}

We produced the NuSTAR images from the cleaned event files using {\tt
  nuproducts}.
In figure~\ref{fig:nuimg} (left panel), we show the
NuSTAR 3--20 keV image.
A significant point source was solely detected within the error circle
(radius $=3\arcmin$) of the MAXI J0709 position 
determined by the NICER observation \citep{ATel15181.I}.
To determine the accurate source position,
we performed image fit 
using {\tt sherpa} included in the Chandra Interactive Analysis of
Observation (ciao; version 4.14).
To improve photon statistics,
images of FPMA and FPMB are combined and binned by 2$\times$2 pixels.
The obtained image in the $4' \times 4'$ region around the count peak was then fitted
with a model consisting of a 2-dimensional (2-D) Gaussian function for the target point source
and a flat surface for the background.
There, we employed the Cash statistics.
Because the ellipticity of the 2-D Gaussian was not significantly detected,
the ellipticity parameter was fixed at 0 (circular).
The best-fit model parameters are the source position ($\alpha$, $\delta$)(J2000)
$=$ (\timeform{7h09m37.1s}, \timeform{-16D05'47''}) with
errors of $2\arcsec$, and the full width at half maximum (FWHM) of
$16\arcsec\pm 2\arcsec$,
where errors include only statistic uncertainties.
The systematic error of the NuSTAR position accuracy
is estimated at $\simeq 20\arcsec$,
which slightly depends on the source brightness \citep{Lansbury2017}.

Comparing the obtained NuSTAR image with 
the DSS (Original Digitized Sky Survey) optical image
provided via the Skyview website\footnote{http://skyview.gsfc.nasa.gov}
in the same region (the right panel of figure
\ref{fig:nuimg}), we identified a possible optical counterpart, LY CMa
(also known as HD 54786), 
which is identified as a Be star \citep{ref:LYCMa_nature}
with a spectral type of B1.5I(b), i.e. B supergiant \citep{1988mcts.book.....H}.
By Gaia Early Data Release 3,
its celestial position 
($\alpha$, $\delta$)(J2000) $=$
(\timeform{7h09m36.9791095248s}, \timeform{-16D05'46.801897476''}), 
and distance $D=3.03^{+0.31}_{-0.27}$ kpc
are well determined
\citep{gaia2016, gaia2020, 2021AJ....161..147B}. 
The position is only \timeform{1.8"} from the NuSTAR best-fit parameters.
The direction and distance suggest that the object locates on the Perseus Arm, a major spiral arm of our Galaxy.

\subsection{Light curve}
\label{sec:nulc}

\begin{figure}
\begin{center}
\includegraphics[width=0.48\textwidth]{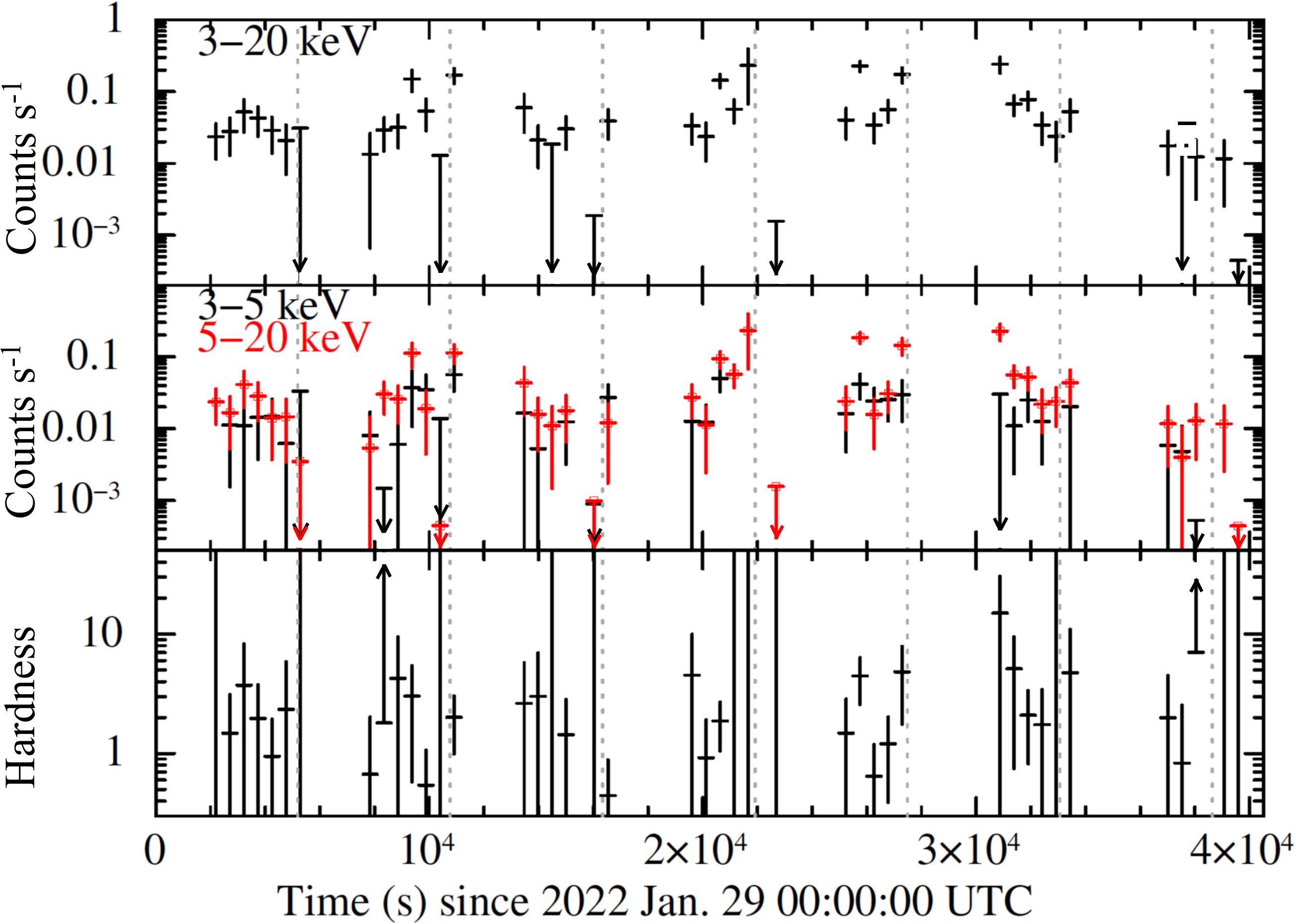}  
\end{center}
\caption{
NuSTAR background-subtracted light curves in 3--20 keV (top), 
3--5 keV (middle; black crosses) and 5--20 keV (middle; red open
squares) with 512 s bins, and the hardness ratio between the latter
two bands (bottom). The background level is 
0.01--0.02 count s$^{-1}$ in the 3--20 keV during the observation.
Vertical dashed lines represent the epochs of MAXI scans 
in figure \ref{fig:maxi_lc}.
} 

\label{fig:nulc}
\end{figure}

We produced the NuSTAR light curves in 3 energy bands,
3--20 keV, 3--5 keV, and 5--20 keV, with {\tt nuproducts}.
Source data were extracted from a circular region with a radius
of $30\arcsec$ centered at the LY CMa position, and 
background data were extracted from 
another circle with a radius of $60\arcsec$ in a source-free region.
In this step, a barycentric time correction
was applied to event files,
assuming the source position of LY CMa.
The data of FPMA and FPMB were
combined with the ftool {\tt lcmath}. 
In figure~\ref{fig:nulc} we plotted the
obtained light curves in the 3 energy bands and
the 5--20 keV to 3--5 keV hardness ratio, with 512 s bins. 
The 3--20 keV light curve shows flux variation by factor of $\lesssim 3$ on timescales of $\gtrsim 10^2$ s.
The middle panel suggests that the variation is larger in the higher
band (i.e., the source becomes harder with increasing flux),
although it is not clear in the hardness-ratio variation.

\subsection{Pulsation search}
\label{sec:pulsearch}
\begin{figure}
\begin{center}
  \includegraphics[width=0.48\textwidth]{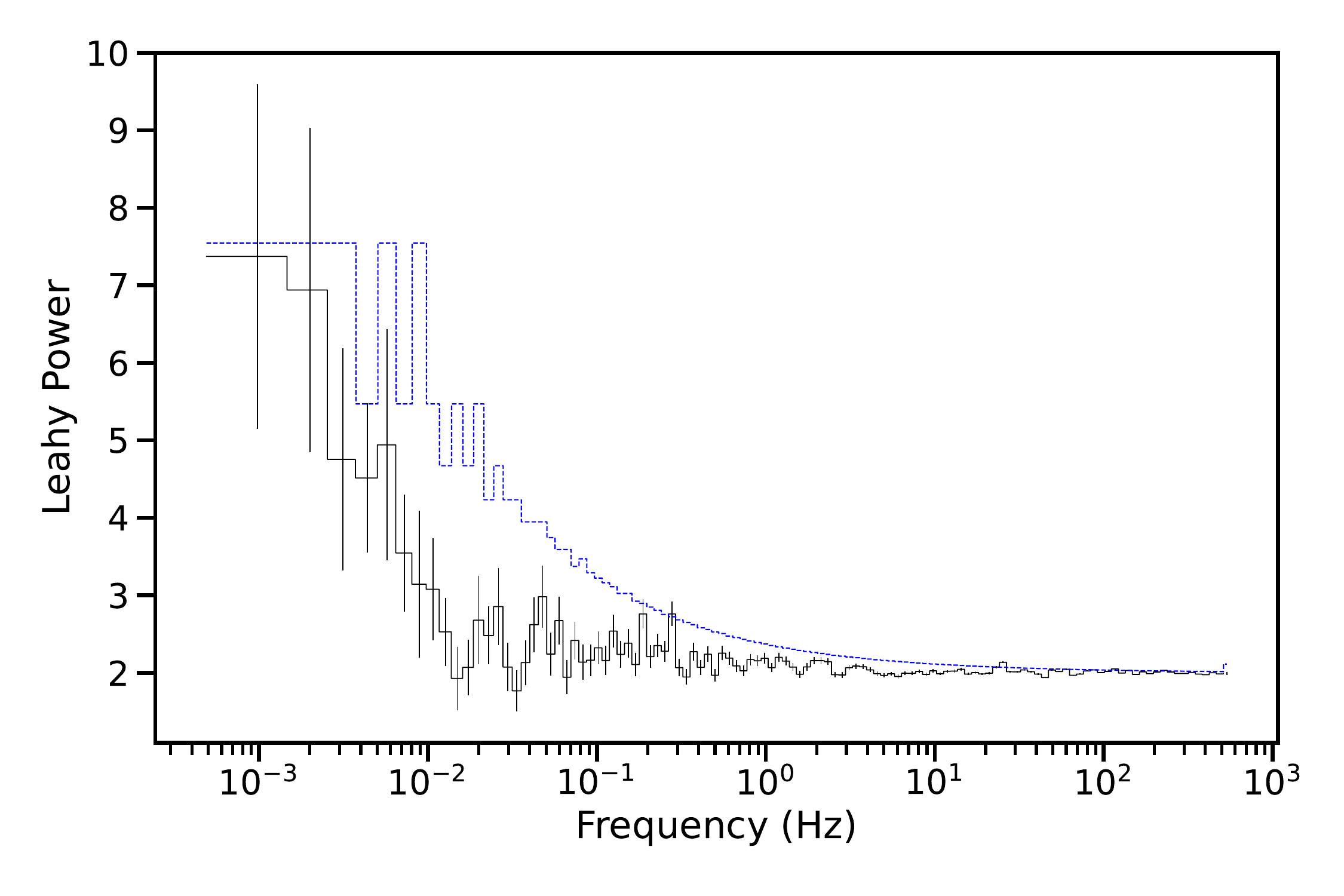}  
\end{center}
\caption{Power spectrum calculated using 3-20\,keV NuSTAR FPMA and
  FPMB data, averaged over the entire observation period. The blue
  dashed line represents the 3$\sigma$ detection level of
  periodicity.}
\label{fig:nupowsp}
\end{figure}

In order to search for coherent pulsations, we produced a power
density spectrum (PDS) from the FPMA$+$FPMB data, which is shown in
figure~\ref{fig:nupowsp}. We first barycenter corrected the time of
arrival for each photon using the source position as determined using
the automatic centroiding function in DS9 \citep{DS9}. We used JPL
ephemeris DE-430 and NuSTAR clockfile v138. We extracted events with
energy 3-20\,keV using a source region with radius
$30\arcsec$. Next we used the Python package Stingray
\citep{Stingray} to produce the Frequency Amplitude
Difference-corrected PDS \citep{Bachetti2018}, which corrects for
timing effects due to deadtime. We specified a time resolution of
$2^{-10}$\,s and a light-curve segment size of 1024\,s, resulting in
13 individual PDS, which we averaged to produce the PDS shown. The PDS
has been rebinned logarithmically for clarity. We also calculated the
$3\sigma$ detection level per logarithmic frequency bin using the
formalism described in \citet{Leahy1983}, which we show as a blue
dashed line.

We did not detect any periodicity significantly, but we note that our
timing analysis is severely limited due to a small number of photons
($<300$ total). As such we cannot rule out the presence of pulsations
or other timing signatures.

\subsection{Time-averaged spectrum}
\label{sec:nuspec}

\begin{figure}
\begin{center}
\includegraphics[width=0.48\textwidth]{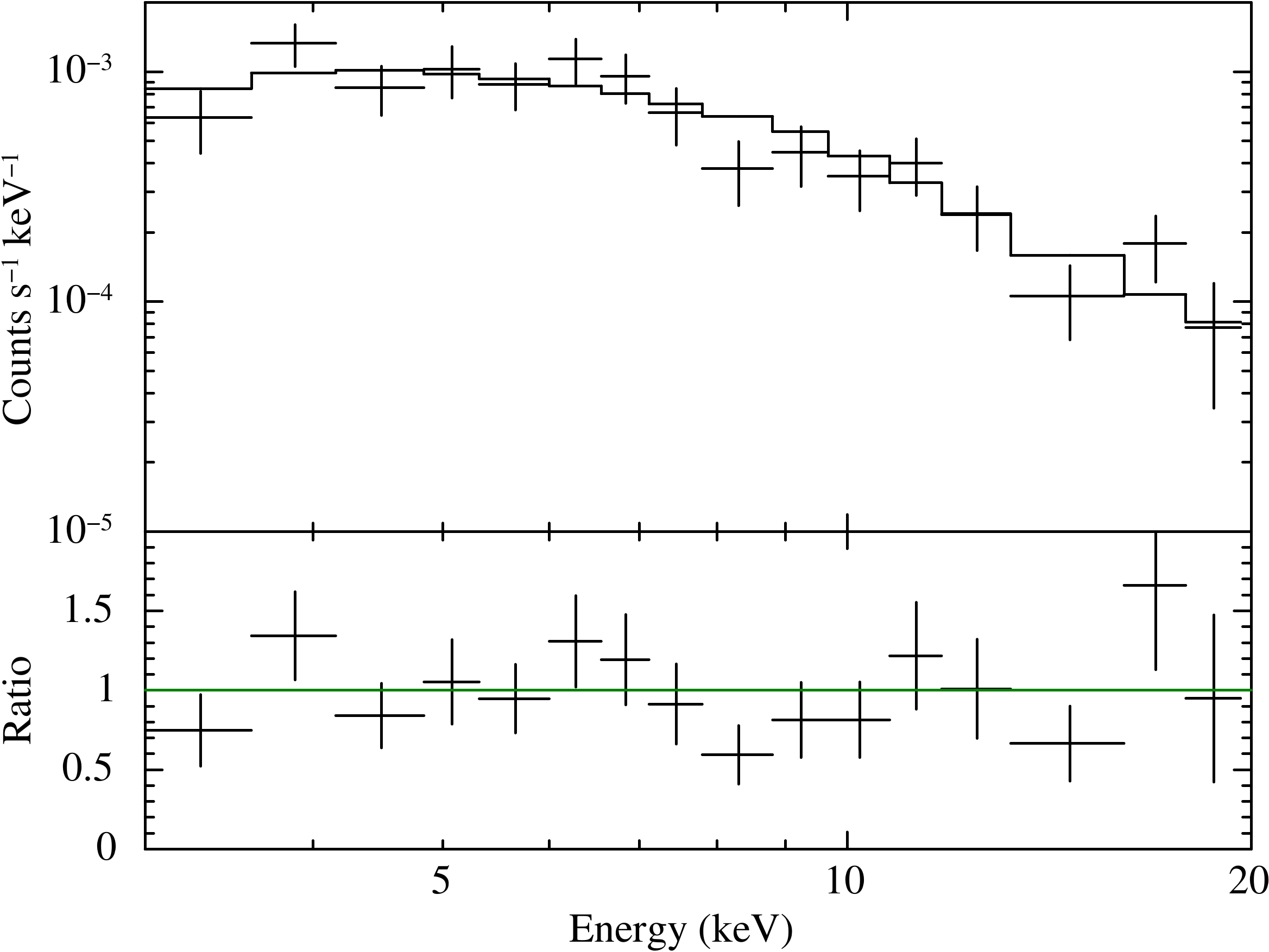}  
\end{center}
\caption{
(Top) Time-averaged NuSTAR/FPMA$+$FPMB
background-subtracted, response-folded spectrum.
The solid line represents the best-fit absorbed power-law model.
(Bottom) Data-to-model ratio. 
Data in this figure were rebinned for visual clarity.
}
\label{fig:nuspec}
\end{figure}

Figure~\ref{fig:nuspec} shows the time-averaged NuSTAR spectrum, which
was extracted with {\tt nuproducts} using the same source and
background regions as in the light curve analysis (section \ref{sec:nulc}).
The FPMA and FPMB spectra were combined with the ftool {\tt addspec}
to improve statistics, and binned so that each bin contains at least one count.
We confirmed that the results did not change 
even if the data of FPMA and FPMB were analyzed separately.

We fitted the obtained spectrum with an absorbed power-law model 
({\tt TBabs*powerlaw}) on XSPEC employing W statistic. 
The results showed the fit statistic $W=148$ for 184 d.o.f.
and the best-fit $\Gamma=1.7^{+0.5}_{-0.2}$. 
$N_\mathrm{H}$ was constrained only by the upper limit, $<1.1 \times 10^{23}$ {\ucolumn},
which is consistent with the result of the MAXI/GSC Scan-A spectrum 
(table \ref{table:gsc_spectra_par} in section \ref{sec:maxispec}). 
From the best-fit model,
the absorbed and absorption-corrected 2--10 keV fluxes 
were estimated to be $6\pm 1 \times 10^{-13}$
and $6^{+3}_{-1} \times 10^{-13}$ 
{\uflux}, respectively.
These spectral parameters are consistent with those
given by \citet{2022arXiv220604473B} which utilized 
the same NuSTAR data.

Although there is no emission-line feature in the NuSTAR spectrum, we
tested the possibility of a narrow iron-K$\alpha$ emission line. 
We added a Gaussian function with a fixed centroid of 6.4 keV and a fixed 
width of 0 eV to the power-law model, and then fit it to the data.
As the result, the 90 \% upper limit on the iron-K$\alpha$ line flux was estimated to be $
4 \times 10^{-6}$ photons cm$^{-2}$ s$^{-1}$, which corresponds to the
equivalent width (EW) of 0.6 keV.

\section{Swift observation and upper limit} 
A ToO observation of the Neil Gehrels Swift Observatory \citep{2004ApJ...611.1005G} was carried out 
on 2022 February 23 (MJD 59634.62), 30 days after the discovery,
with a total exposure time of 989 s. 
No X-ray source was detected with the $3\sigma$ upper limit  
of 0.008 XRT counts s$^{-1}$ at the position of LM CMa.
Assuming spectral parameters given in table \ref{table:gsc_spectra_par}, 
this provides an upper limit on the flux of $2\times10^{-12}$ {\uflux} (0.3--10 keV) for the Scan-A parameters, 
and $1.3\times10^{-12}$ {\uflux} (0.3--10 keV) for the Scan-B parameters, 
where the flux is not corrected for absorption. 
These upper limits are consistent with the detected flux by NuSTAR.

\section{eROSITA upper limits on the past activity} 
\label{sec:erosita}
The {\it Spectrum-Roentgen-Gamma (Spektr-RG, SRG)} X-ray observatory
\citep{2021A&A...656A.132S}
has been performing an all-sky survey 
since 2019 December. 
The satellite carries two kinds of X-ray telescopes, extended ROentogen Survey with 
an Imaging Telescope Array (eROSITA; \cite{2021A&A...647A...1P}),
and Mikhail Pavllinsky Astronomical Roentgen Telescope (ART-XC; \cite{2021A&A...650A..42P}),
which cover 0.2--8 keV and 4--30 keV energy bands, respectively.
The eROSITA all-sky survey (eRASS) has been carried out by consecutive
scans of the entire sky, each of which is completed by six months.
It currently archives the best sensitivity among all-sky X-ray surveys
that have ever been performed.

The sky position of MAXI J0709 was covered by eROSITA in the past four eRASSs
as listed in Table \ref{tab:eROSITA_obs}. 
All of eROSITA data are calibrated and cleaned using the pipeline version 946 of the eROSITA Science Analysis Software System
(eSASS, \cite{2022A&A...661A...1B}). 
No significant X-ray emission at the position of LY CMa was detected in either the individual eRASS or the combined dataset. 
We estimated the 3-$\sigma$ upper limit of $4\times10^{-13}$ {\uflux}
in 2--10 keV band using all the eRASS data 
and assuming the typical power-law spectrum
of $\Gamma=2$ and the interstellar absorption corresponding to
the Galactic $\mathrm{H}_\mathrm{I}$ density $N_\mathrm{H}=0.5\times 10^{22}$ {\ucolumn}.

\begin{table}
\begin{threeparttable}
\caption{eRASS observations on the MAXI J0709 position (upper limits).}
\label{tab:eROSITA_obs}
\begin{tabular}{cccc}
\hline
\hline
\#$^{a}$  & Start time & End time & $T_\mathrm{exp}^{b}$\\
\hline
1 & 2020-04-17 09:34:14 & 2020-04-18 01:34:44 & 173\\
2 & 2020-10-20 08:25:17 & 2020-10-21 00:25:41 & 156\\
3 & 2021-04-18 02:34:27 & 2021-04-19 06:34:33 & 234\\ 
4 & 2021-10-20 13:25:24 & 2021-10-21 13:25:39 & 219\\
\hline
\end{tabular}
\begin{tablenotes}
\item[a] eRASS survey number.
\item[b] Exposure time (s).
\end{tablenotes}
\end{threeparttable}
\end{table}

\section{Optical spectroscopy with SCAT} 
\label{sec:scat}
The Spectroscopic Chuo university Astronomical Telescope (SCAT) is a 355-mm diameter optical telescope 
equipped with an ATIK 460EX CCD camera and a Shelyak Alpy 600 spectrometer,
located at the Chuo Univeristy Korakuen campus, Tokyo, Japan \citep{2022PASJ...74..477K}. 
The spectral resolution is R$\sim$600.

\begin{figure}
\begin{center}
\includegraphics[angle=270,width=0.48\textwidth]{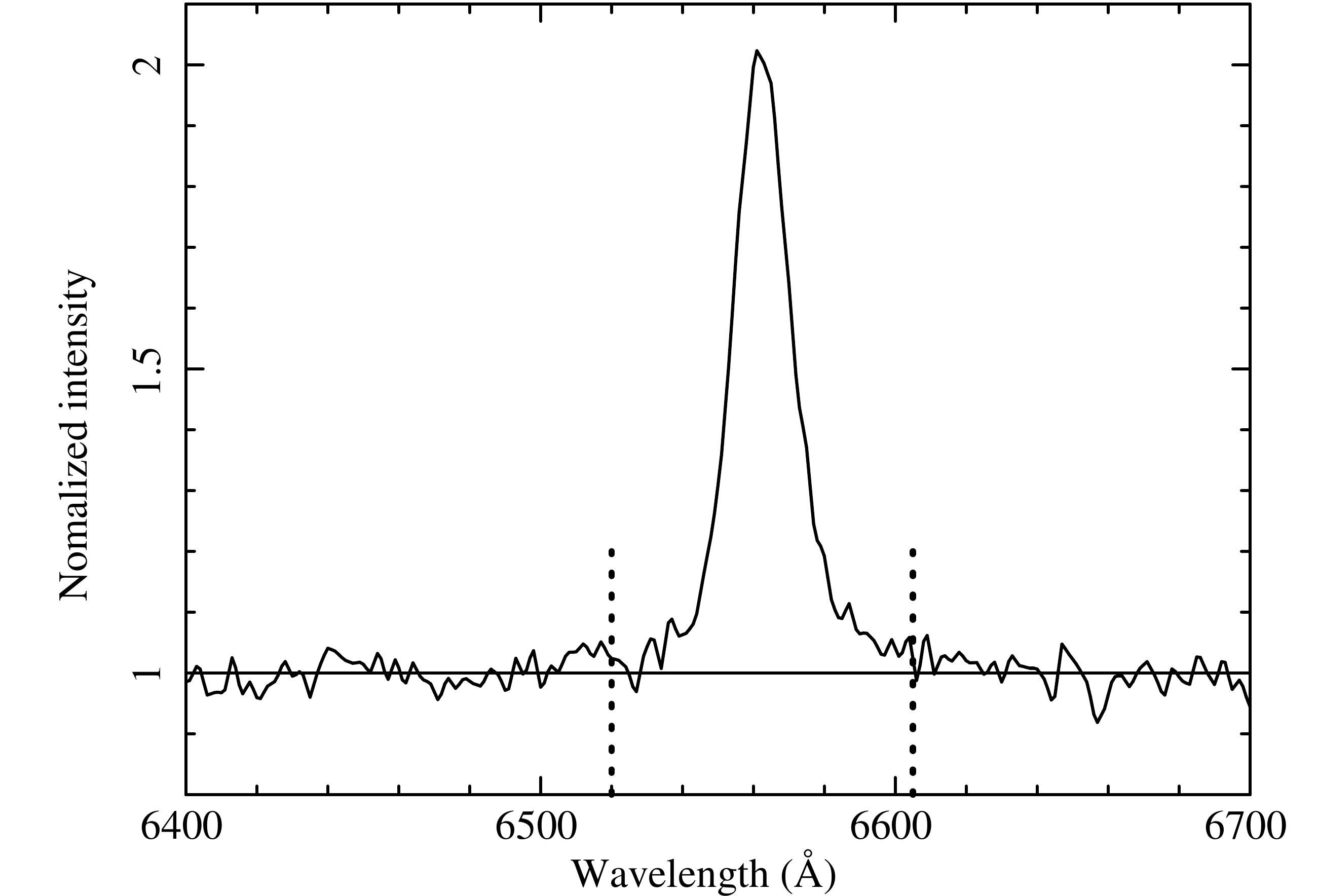}  
\end{center}
\caption{
Optical spectrum of LY CMa obtained by SCAT around the H$\alpha$ emission line.
The intensity scale is normalized by the linear continuum model.
The region between two dashed lines represents the range of the H$\alpha$ line employed in the EW estimate.
}
\label{fig:scatspec}
\end{figure}

The SCAT observation of LY CMa was carried out on 2022 January 28 from UT 11:32 to 14:48 
(approximately 3 days after the MAXI trigger) with a net exposure of 8610 s. 
The result showed a strong H$\alpha$ emission line clearly. 
The FWHM of the H$\alpha$ line was 19 {\AA}, 
which is
significantly larger than that of the Ne line in the calibration lamp
(FWHM$=10$ {\AA}). 
Next, we estimated the line EW.
To determine the continuum level,
we fitted the observed spectrum excluding the H$\alpha$-line range
with a linear function.
Figure \ref{fig:scatspec} shows the line profile normalized by the best-fit continuum model,
where the H$\alpha$-line range was assumed to be from 6520 {\AA} to 6605 {\AA}.
By integrating the profile, the EW was estimated to be $-23$ {\AA}.  
It is consistent with the result of the Foligno Observatory low-resolution (R$\sim$50) spectrum taken
on the same day \citep{ATel15194.N}.
We repeated the EW calculation by shifting the assumed H$\alpha$-line range, and found that the obtained EW value has an uncertainty of $\sim 15$\%.

\section{Discussion} 

\subsection{Summary of MAXI J0709-159 activities with past and follow-up observations}

We here summarize long-term activities of MAXI J0709 
identified with the past and follow-up observations in the present analysis.
Based on the results, we consider
the nature of the new X-ray object in the next section.

The new transient MAXI J0709 was first discovered by the MAXI GSC
all-sky survey on 2022 January 25.
As seen in figure \ref{fig:maxi_lc},
significant X-rays were detected only in the
two GSC scans, Scan-A (UT 10:42) and Scan-C (UT 14:48).
In both scans, it showed
flare-like time variabilities in a time scale of a few seconds (figure \ref{fig:scan_lc})
and the 2--10 keV flux reached $5\times 10^{-9}$ {\uflux}.
At Scan-B (UT 12:15) and after the Scan-C (UT 15:21), 
the source has not been detected with the GSC sensitivity limit
of 80 mCrab ($1.5\times 10^{-9}$ {\uflux} in 2--10 keV) per scan
\citep{Negoro2016PASJ}.
Therefore, the intensity swings between every adjacent scans from Scan-A to Scan-D
exceed a factor of $\sim 5$.

The NICER and NuSTAR follow-up observations
successfully identified the new transient with a new X-ray source, 
whose position is consistent LY CMa
(figure \ref{fig:nuimg}).
When the NICER observed it at 6 minutes after the MAXI Scan-C,
the X-ray flux became about 10 mCrab (in 0.2-12 keV),
which is $< 10^{-1}$ of that observed in Scan-C.
When the NuSTAR identified the new object on January 29,
4 days after the MAXI detections,
the X-ray flux became relatively stable at around $6\times 10^{-13}$ erg cm$^{-2}$ s$^{-1}$.
This means that the X-ray intensity decreased by a factor of $10^{-4}$ in the 3 days.

We investigated the past source activity using archival data.
Until the first MAXI detection,
the source had not been recognized in the MAXI GSC data for over 12 years.
The upper limit on the average source flux is $5\times 10^{-12}$ {\uflux}
\citep{2018ApJS..235....7H}.
The object had not been recorded in any X-ray source catalog
including ROSAT all-sky survey catalog \citep{2016A&A...588A.103B}.
%
Also, eROSITA all-sky surveys which observed the MAXI J0709 position four times from 2020 to 2021,
did not find significant X-ray source
with the 3-$\sigma$ upper limit of $4\times 10^{-13}$ {\uflux} (section \ref{sec:erosita}),
which is lower than the flux observed by NuSTAR.
This suggests that the source activity at the NuSTAR observation
was still higher than that before the discovery.

\begin{figure*}
\begin{center}
\includegraphics[width=0.75\textwidth]{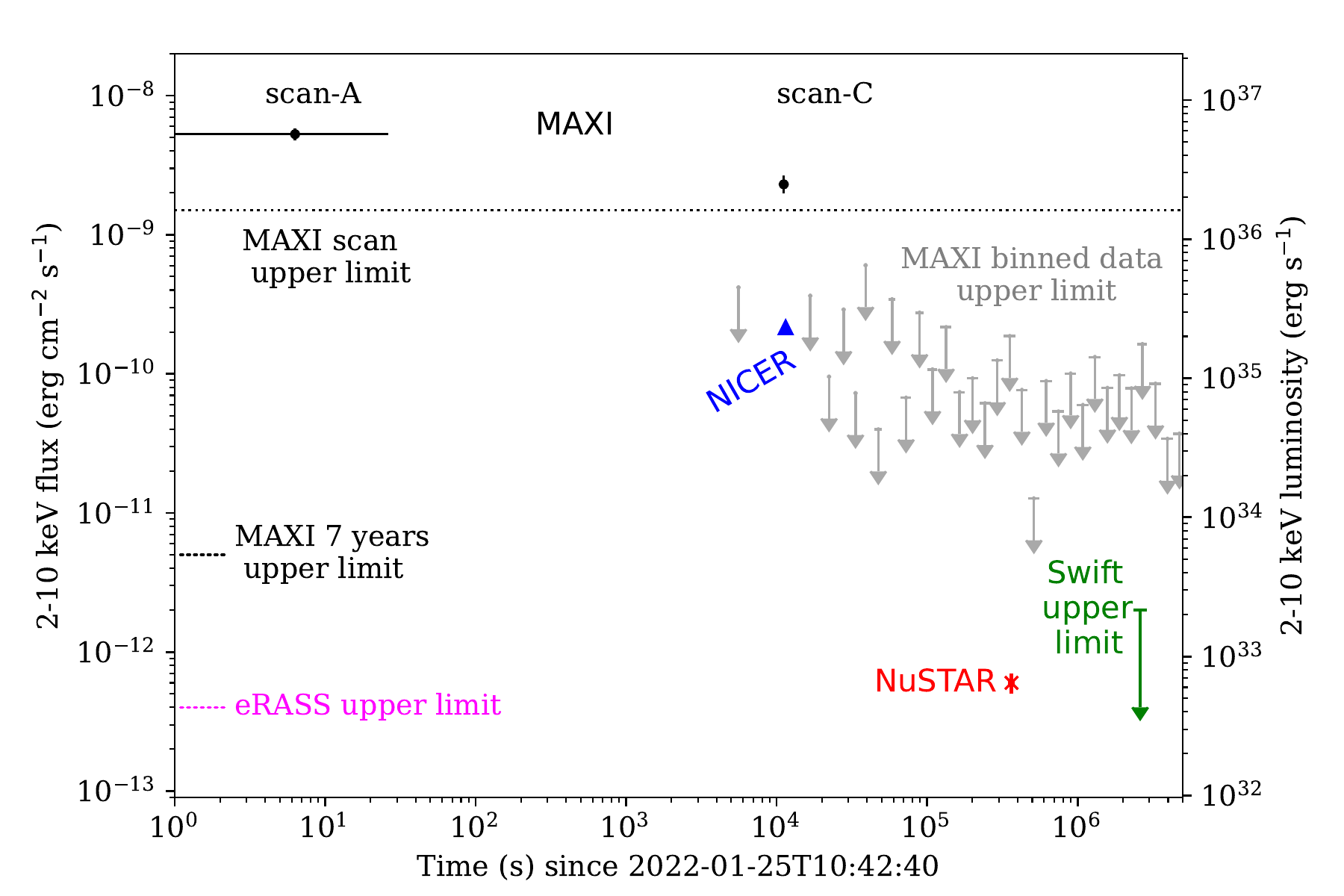}
\end{center}
\caption{
  MAXI J0709 long-term X-ray activity since the first MAXI detection at UT 10:42 on 2022 January 25,
  obtained from the present analysis of the MAXI, NuSTAR, Swift, and eROSITA data and the NICER result
  reported by \citet{ATel15181.I}.
  Upper limits from the logarithmically-binned MAXI/GSC light curve 
  are shown in gray arrows.
}
\label{fig:lcsum}
\end{figure*}

In figure \ref{fig:lcsum}, 
the long-term activities after the discovery of MAXI J0709
are summarized.
The X-ray intensity variation clearly reveals that the activity declined at $\sim 10^{4}$ s ($\simeq 3$ hours).  
In the figure, the right-hand ordinate represents
the luminosity calculated from the flux at the source distance of 3 kpc.
The observed flux range of $10^{-13}$--$10^{-8}$ {\uflux}
corresponds to $10^{32}$--$10^{37}$ {\ulumi} in the luminosity.

We also carried out an optical spectroscopic observation
of the optical counterpart LY CMa on January 28, 3 days after the discovery,
and then confirmed the H$\alpha$ emission line
(section \ref{sec:scat})
as reported in \citet{ATel15194.N}.
The EW of the H$\alpha$ line was estimated to be 
$-23$ {\AA} with an uncertainty of $\sim 15$\,\%.
\citet{2022arXiv220604473B} 
also performed optical follow-up observations and 
reported that  
the H$\alpha$ EW was
$-17.6$ {\AA} on February 2 
and $-16.9$ {\AA} on February 3.
The results suggest that the EW might change in these 3 days.

\subsection{Classification of the new X-ray object in high-mass X-ray binaries}

As discussed in the previous section and summarized in figure \ref{fig:lcsum},
the observed X-ray activity of MAXI J0709 is characterized by 
the initial flaring phase of $\approx 3$ hours
with a rapid ($\sim$ few seconds) variability reaching the peak luminosity $\simeq 5\times 10^{36}$ {\ulumi},
and the following decay phase lasting for at least several days with a luminosity
$\sim 10^{32}$--$10^{33}$ {\ulumi} and a moderate variation.
The rapid variability naturally lead us an idea of X-ray binaries
embedding compact objects, neutron stars or black holes, where
the transient behavior can be explained by the change
in the mass accretion onto the compact object. 
In the present MAXI J0709 case, the mass-donating stellar companion is
identified with LY CMa, which has a spectral type of B1.5I(b) \citep{1988mcts.book.....H}
and also the identification as Be star \citep{ref:LYCMa_nature}.
Hence, we consider the detail scenario 
to explain all the observed results
in comparison with other high-mass X-ray binaries (HMXBs) that have been well studied.

The initial flaring behavior of the short duration ($\lesssim$ several hours) and
the rapid variability (in a time scale of a few seconds),
agree well with the properties of Supergiant Fast X-ray Transients (SFXTs),
a possible subclass of HMXBs that consist of OB supergiants and neutron stars
showing sporadic short-duration ($\lesssim$ several hours) outbursts
(\cite{2006ApJ...646..452S,2015AdSpR..55.1255B,2018MNRAS.481.2779S}; review in  \cite{2019NewAR..8601546K}).
The observed luminosity range from $10^{32}$ to $10^{37}$ {\ulumi}
agree with those from their quiescent value to the flaring peak.
The X-ray spectra fitted with a power law of $\Gamma\simeq 2$
in both MAXI and NuSTAR data 
are also typical as SFXTs which have power-law spectra with $\Gamma\sim 1(\pm1)$ \citep{2018A&A...610A..50P}.

During the initial flaring phase,
the X-ray spectrum changed dramatically so that $N_\mathrm{H}$ changed from $10^{22}$ to $10^{23}$ {\ucolumn} in $\sim 3$ hours
(figure \ref{fig:koba_gsc_spec}).
The $N_\mathrm{H}\gtrsim 10^{23}$ {\ucolumn} at the MAXI GSC Scan-C
is significantly higher than the Galactic $\mathrm{H}_\mathrm{I}$ density $\simeq 0.5\times 10^{22}$ cm$^{-2}$,
and also comparable to the highest among those of SFXTs that have ever been reported
\citep{2018A&A...610A..50P}.
So far, a similar spectral change has been observed in IGR J18410-0535, and it was explained by a scenario that 
the compact object (neutron star) just plunged into
a dense clump in the circumstellar medium
\citep{2011A&A...531A.130B}.
The present result can be considered similarly.

Also, iron-K line was not significantly detected in either MAXI or NuSTAR spectrum.
From the NuSTAR data, the upper limit on the equivalent width $EW_\mathrm{FeK}<0.6$ keV was estimated.
In general, X-ray spectra of the SFXTs
have lower $N_\mathrm{H}$ and lower $EW_\mathrm{FeK}$
than those of the classical supergiant HMXBs
with persistent X-ray activities \citep{2015A&A...576A.108G,2018A&A...610A..50P}.
This implies that the SFXTs have more sparse and clumpy circumstellar media than the classical supergiant HMXBs.
The present results support the scenario.

To explain the short duration (several hours)
and the extremely large dynamic range ($10^{3}$--$10^{4}$) from the quiescence to the peak
in the SFXT outbursts,
some mechanism to inhibit accretion
such as magnetic and/or centrifugal barriers
\citep{2007AstL...33..149G,2008ApJ...683.1031B}
are required.
%
If compact objects in SFXTs are magnetized neutron stars,
observed X-ray variations may relate to
the neutron star rotation.
In fact, coherent pulsations were detected in several SFXTs.
We performed the period search with the NuSTAR data,
but could not find a significant periodicity (section \ref{sec:pulsearch}).
The results are consistent with the most of major well-known SFXTs,
on which the periodicity have not been detected
even though deep observations were carried out
(e.g. \cite{2010A&A...519A...6B,2020A&A...642A..73F}).
This may indicate that the neutron stars have very long spin periods
and the X-ray variations only comes form the magnetic and centrifugal
gating mechanism \citep{2008ApJ...683.1031B}, or the accretions are in
a quasi-spherical accretion regime \citep{2012MNRAS.420..216S}.

\subsection{Is MAXI J0709-159 an evolved Be Fast X-ray Transient ?}

So far, we have discussed the nature of MAXI J0709 based on the X-ray
results that agree well with the typical SFXTs.
Meanwhile, the optical counterpart LY CMa is also identified as Be star \citep{ref:LYCMa_nature}.
In fact, optical follow-up observations confirmed
the broad H$\alpha$ line, which indicates
the Be circumstellar disk 
(section \ref{sec:scat}). 
\citet{2022arXiv220604473B} proposed that
LY CMa would be an evolved Be star
located between main sequence stars and supergiants
on the optical color-magnitude diagram.
If it is so, MAXI J0709 should
have an intermediate character between
Be X-ray binaries (BeXBs) and supergiant X-ray binaries (sgXBs).

BeXBs usually cause outbursts
lasting for several weeks (type-I) to a few months (type-II),
according to the interaction between the neutron-star magnetosphere and the Be disk
(e.g. \cite{Reig2011, 2017PASJ...69..100S}).
The behavior is quite different from the present MAXI J0709 results.
However, a few BeXBs, such as X Persei,
are known to show short time-scale variabilities like flares \citep{2001ApJ...546..455D,2014MNRAS.444..457A}.
MAXI J0709 could be considered as its extreme case.

Recently, another new subclass of HMXBs involving supergiant B[e]
(sgB[e]) stars 
as mass-donating companions, is getting a hot topic
(e.g. \cite{2019Galax...7...83K}).
The sgB[e] stars are thought to accompany dense, dusty disks.
The two members of this subclass, CI Cam \citep{2019A&A...622A..93B}
and IGR J16318$-$4848 \citep{2020ApJ...894...86F}, 
are known to exhibit extremely high
($\gtrsim 10^{24}$ {\ucolumn}) and variable $N_\mathrm{H}$ on their
persistent X-ray activities.
Although LY CMa is not definitely categorized into sgB[e],
the thick and variable X-ray absorption feature in these two
objects is quite similar with that observed in MAXI J0709
(section \ref{sec:maxispec}).
This naturally induces an idea that 
some sgB[e] HMXBs may behave like SFXT.
A candidate of such objects, namely sgB[e]FXTs, has been already reported \citep{2022MNRAS.512.2929S}.
Further observations of LY CMa
would give us useful hints about possible relations among these HMXB subclasses.

\section{Conclusion}

On 2022 January 25, MAXI discovered the new bright X-ray transient MAXI J0709-159 
lasting for $\sim 3$ hours 
in the constellation Canis Majoris. 
Prompt follow-up observations with NICER and NuSTAR confirmed
the new X-ray object and refined the source position with the $20\arcsec$ accuracy.
Then, the optical counterpart was identified with 
LY CMa, which has been identified as
B supergiant and also Be star. 
Detail analysis of MAXI and NuSTAR data revealed the characteristic X-ray outburst 
represented by the short duration ($\sim 3$ hours),
the rapid ($\lesssim$ a few seconds) variability 
accompanied with the spectral change, 
and large luminosity swing from the quiescence ($\sim10^{32}$ {\ulumi}) to the flare peak  ($\sim 10^{37}$ {\ulumi}).
These features agree well with the typical SFXTs.
The spectral change during the short outburst period 
suggests that the $N_\mathrm{H}$ increased 
from $10^{22}$ to $10^{23}$ {\ucolumn},
which can be explained by a scenario that 
the compact object (neutron star) just plunged into
a dense clump in the circumstellar medium.
Meanwhile, the optical spectroscopic observation of LY CMa 
reveals the broad H$\alpha$ emission line suggesting
the existence of the circumstellar Be disk.
However, the observed X-ray behavior agrees with the SFXT, i.e.
supergiant X-ray binaries rather than Be X-ray binaries.
Thus, LY CMa is surrounded by  
complex circumstellar medium including dense clumps.
These facts suggests that the object would be classified into
the intermediate position between
these HMXB subclasses, namely
evolved Be Fast X-ray Transient.

\begin{ack}
The authors thank the MAXI team members for their
dedicated work on the mission operation.
M. Sugizaki acknowledges support from the Chinese Academy of
Sciences (CAS) President's International Fellowship Initiative (PIFI)
(grant No. 2020FSM004).
Part of this work was financially supported 
by Grants-in-Aid for Scientific Research 19K14762 (M. Shidatsu) and
21K03620 (H. Negoro)
from the Ministry of Education, Culture, Sports, Science and 
Technology (MEXT) of Japan. 
We thank the NuSTAR and Swift operation teams 
for performing the ToO observations, and Brian Grefenstette for 
the quick look analysis of the NuSTAR data. 
%
%
This work is based on data from eROSITA, the soft X-ray instrument aboard SRG, a joint Russian-German science mission supported by the Russian Space Agency (Roskosmos), in the interests of the Russian Academy of Sciences represented by its Space Research Institute (IKI), and the Deutsches Zentrum f{\"u}r Luft- und Raumfahrt (DLR). The SRG spacecraft was built by Lavochkin Association (NPOL) and its subcontractors, and is operated by NPOL with support from the Max Planck Institute for Extraterrestrial Physics (MPE). The development and construction of the eROSITA X-ray instrument was led by MPE, with contributions from the Dr. Karl Remeis Observatory Bamberg \& ECAP (FAU Erlangen-Nuernberg), the University of Hamburg Observatory, the Leibniz Institute for Astrophysics Potsdam (AIP), and the Institute for Astronomy and Astrophysics of the University of Tübingen, with the support of DLR and the Max Planck Society. The Argelander Institute for Astronomy of the University of Bonn and the Ludwig Maximilians Universität Munich also participated in the science preparation for eROSITA.
%
%
This work has also made use of data from the European Space Agency (ESA) mission
{\it Gaia} (\url{https://www.cosmos.esa.int/gaia}), processed by the {\it Gaia}
Data Processing and Analysis Consortium (DPAC,
\url{https://www.cosmos.esa.int/web/gaia/dpac/consortium}). Funding for the DPAC
has been provided by national institutions, in particular the institutions
participating in the {\it Gaia} Multilateral Agreement.
%
%
The Digitized Sky Surveys were produced at the Space Telescope Science
Institute under U.S. Government grant NAG W-2166. The images of these
surveys are based on photographic data obtained using the Oschin
Schmidt Telescope on Palomar Mountain and the UK Schmidt
Telescope. The plates were processed into the present compressed
digital form with the permission of these institutions.  The National
Geographic Society - Palomar Observatory Sky Atlas (POSS-I) was made
by the California Institute of Technology with grants from the
National Geographic Society.  The Second Palomar Observatory Sky
Survey (POSS-II) was made by the California Institute of Technology
with funds from the National Science Foundation, the National
Geographic Society, the Sloan Foundation, the Samuel Oschin
Foundation, and the Eastman Kodak Corporation.  The Oschin Schmidt
Telescope is operated by the California Institute of Technology and
Palomar Observatory.  The UK Schmidt Telescope was operated by the
Royal Observatory Edinburgh, with funding from the UK Science and
Engineering Research Council (later the UK Particle Physics and
Astronomy Research Council), until 1988 June, and thereafter by the
Anglo-Australian Observatory. The blue plates of the southern Sky
Atlas and its Equatorial Extension (together known as the SERC-J), as
well as the Equatorial Red (ER), and the Second Epoch [red] Survey
(SES) were all taken with the UK Schmidt.  Supplemental funding for
sky-survey work at the ST ScI is provided by the European Southern
Observatory.
\end{ack}


\begin{thebibliography}{}


\bibitem[Acuner et al.(2014)]{2014MNRAS.444..457A} Acuner, Z., {\.I}nam, S. {\c{C}}., {\c{S}}ahiner, {\c{S}}., et al.\ 2014, \mnras, 444, 457. doi:10.1093/mnras/stu1351

\bibitem[Arnaud (1996)]{Arnaud1996}
Arnaud, K. ~A.\ 1996, in ASP Conf. Ser., 101, 
Astronomical Data Analysis Software and Systems V, ed. G. H. Jacoby \& J. Barnes (San Francisco, CA: ASP), 17

\bibitem[Bachetti \& Huppenkothen(2018)]{Bachetti2018} Bachetti, M. \& Huppenkothen, D.\ 2018, \apjl, 853, L21. doi:10.3847/2041-8213/aaa83b

\bibitem[Bailer-Jones et al.(2021)]{2021AJ....161..147B} Bailer-Jones, C.~A.~L., Rybizki, J., Fouesneau, M., et al.\ 2021, \aj, 161, 147. doi:10.3847/1538-3881/abd806

\bibitem[Bartlett et al.(2019)]{2019A&A...622A..93B} Bartlett, E.~S., Clark, J.~S., \& Negueruela, I.\ 2019, \aap, 622, A93. doi:10.1051/0004-6361/201834315

\bibitem[Bhattacharyya et al.(2022)]{2022arXiv220604473B} Bhattacharyya, S., Mathew, B., Ezhikode, S.~H., et al.\ 2022, arXiv:2206.04473


\bibitem[Boller et al.(2016)]{2016A&A...588A.103B} Boller, T., Freyberg, M.~J., Tr{\"u}mper, J., et al.\ 2016, \aap, 588, A103. doi:10.1051/0004-6361/201525648

\bibitem[Bozzo et al.(2008)]{2008ApJ...683.1031B} Bozzo, E., Falanga, M., \& Stella, L.\ 2008, \apj, 683, 1031. doi:10.1086/589990

\bibitem[Bozzo et al.(2010)]{2010A&A...519A...6B} Bozzo, E., Stella, L., Ferrigno, C., et al.\ 2010, \aap, 519, A6. doi:10.1051/0004-6361/201014095

\bibitem[Bozzo et al.(2011)]{2011A&A...531A.130B} Bozzo, E., Giunta, A., Cusumano, G., et al.\ 2011, \aap, 531, A130. doi:10.1051/0004-6361/201116726

\bibitem[Bozzo et al.(2015)]{2015AdSpR..55.1255B} Bozzo, E., Romano, P., Ducci, L., et al.\ 2015, Advances in Space Research, 55, 1255. doi:10.1016/j.asr.2014.11.012

\bibitem[Brunner et al.(2022)]{2022A&A...661A...1B} Brunner, H., Liu, T., Lamer, G., et al.\ 2022, \aap, 661, A1. doi:10.1051/0004-6361/202141266


\bibitem[Cash (1979)]{Cash1979}
Cash, W.\ 1979, \apj, 228, 939

\bibitem[Chojnowski et al.(2015)]{ref:LYCMa_nature} Chojnowski, S.~D.,  Whelan, D.~G., Wisniewski, J.~P., et al. 2015, \aj, 149, 7

\bibitem[Delgado-Mart{\'\i} et al.(2001)]{2001ApJ...546..455D} Delgado-Mart{\'\i}, H., Levine, A.~M., Pfahl, E., et al.\ 2001, \apj, 546, 455. doi:10.1086/318236


\bibitem[Ferrigno et al.(2020)]{2020A&A...642A..73F} Ferrigno, C., Bozzo, E., \& Romano, P.\ 2020, \aap, 642, A73. doi:10.1051/0004-6361/202038278

\bibitem[Fortin et al.(2020)]{2020ApJ...894...86F} Fortin, F., Chaty, S., \& Sander, A.\ 2020, \apj, 894, 86. doi:10.3847/1538-4357/ab881c
 
\bibitem[Gaia Collaboration et al.(2016)]{gaia2016}
Gaia Collaboration et al., 2016, \aap, 595, 1

\bibitem[Gaia Collaboration et al.(2020)]{gaia2020}
Gaia Collaboration et al., 2020, \aap, 649, 1


\bibitem[Gehrels et al.(2004)]{2004ApJ...611.1005G} Gehrels, N., Chincarini, G., Giommi, P., et al.\ 2004, \apj, 611, 1005. doi:10.1086/422091


\bibitem[Gendreau et al.(2012)]{Gendreau2012} Gendreau, K.~C., Arzoumanian, Z., \& Okajima, T.\ 2012, Proc. SPIE, 8443, 844313

\bibitem[Gim{\'e}nez-Garc{\'\i}a et al.(2015)]{2015A&A...576A.108G} Gim{\'e}nez-Garc{\'\i}a, A., Torrej{\'o}n, J.~M., Eikmann, W., et al.\ 2015, \aap, 576, A108. doi:10.1051/0004-6361/201425004

\bibitem[Grebenev \& Sunyaev(2007)]{2007AstL...33..149G} Grebenev, S.~A. \& Sunyaev, R.~A.\ 2007, Astronomy Letters, 33, 149. doi:10.1134/S1063773707030024


\bibitem[Harrison et al.(2013)]{Harrison2013} Harrison, F.~A., Craig, W.~W., Christensen, F.~E., et al.\ 2013, \apj, 770, 103


\bibitem[HI4PI Collaboration et al.(2016)]{2016A&A...594A.116H} HI4PI Collaboration, Ben Bekhti, N., Flöer, L., et al.\ 2016, \aap, 594, A116. doi:10.1051/0004-6361/201629178


\bibitem[Houk \& Smith-Moore(1988)]{1988mcts.book.....H} Houk, N. \& Smith-Moore, M.\ 1988, Michigan Catalogue of Two-dimensional Spectral Types for the HD Stars. Volume 4, Declinations -26{\textdegree}.0 to -12{\textdegree}.0.. N. Houk, M. Smith-Moore.Department of Astronomy, University of Michigan, Ann Arbor, MI 48109-1090, USA. 14+505 pp. Price US 25.00 (USA, Canada), US 28.00 (Foreign) (1988).
  

\bibitem[Hori et al.(2018)]{2018ApJS..235....7H} Hori, T., Shidatsu, M., Ueda, Y., et al.\ 2018, \apjs, 235, 7. doi:10.3847/1538-4365/aaa89c

\bibitem[Huppenkothen et al.(2019)]{Stingray} Huppenkothen, D., Bachetti, M., Stevens, A.~L., et al.\ 2019, \apj, 881, 39. doi:10.3847/1538-4357/ab258d

\bibitem[Iwakiri et al.(2022)]{ATel15181.I} Iwakiri, W., Gendreau, K., Arzoumanian, Z., et al.\ 2022, The Astronomer's Telegram, 15181

\bibitem[Joye \& Mandel(2003)]{DS9} Joye, W.~A. \& Mandel, E.\ 2003, Astronomical Data Analysis Software and Systems XII, 295, 489

\bibitem[Kawai et al.(2022)]{2022PASJ...74..477K} Kawai, H., Tsuboi, Y., Iwakiri, W.~B., et al.\ 2022, \pasj, 74, 477. doi:10.1093/pasj/psac008

\bibitem[Kobayashi et al.(2022)]{ATel15188.K} Kobayashi, K., Negoro, H., Sugita, S., et al.\ 2022, The Astronomer's Telegram, 15188

\bibitem[Kraus(2019)]{2019Galax...7...83K} Kraus, M.\ 2019, Galaxies, 7, 83. doi:10.3390/galaxies7040083

\bibitem[Kretschmar et al.(2019)]{2019NewAR..8601546K} Kretschmar, P., F{\"u}rst, F., Sidoli, L., et al.\ 2019, \nar, 86, 101546. doi:10.1016/j.newar.2020.101546


\bibitem[Lansbury et al.(2017)]{Lansbury2017} Lansbury, G.~B., Stern, D., Aird, J., et al.\ 2017, \apj, 836, 99 

\bibitem[Leahy et al.(1983)]{Leahy1983} Leahy, D.~A., Darbro, W., Elsner, R.~F., et al.\ 1983, \apj, 266, 160. doi:10.1086/160766



\bibitem[Matsuoka et al. (2009)]{Matsuoka2009}
Matsuoka, M., et al.2009, \pasj, 61, 999

\bibitem[Mihara et al. (2011)]{Mihara2011}
Mihara, T., et al. 2011, \pasj, 63, S623

\bibitem[Mihara et al. (2022)]{Mihara2022Handbook}
Mihara, T., Tsunemi, H. \& Negoro, H.\ 2022, "MAXI: Monitor of All-sky X-ray Image" in Handbook of X-ray and Gamma-ray Astrophysics
ed. C. Bambi \& A. Santangelo, arXiv:2206.01505


\bibitem[Morii et al.(2013)]{2013ApJ...779..118M} Morii, M., Tomida, H., Kimura, M., et al.\ 2013, \apj, 779, 118. doi:10.1088/0004-637X/779/2/118

\bibitem[Morii et al.(2016)]{Morii2016} 
Morii, M., Yamaoka, H., Mihara, T., et al.\ 2016, \pasj, 68, S11. doi:10.1093/pasj/psw007


\bibitem[Negoro et al.(2016)]{Negoro2016PASJ} Negoro, H., Kohama, M., Serino, M., et al.\ 2016, \pasj, 68, S1. doi:10.1093/pasj/psw016

\bibitem[Negoro et al.(2022)]{ATel15193.N} Negoro, H., Mihara, T., Pike, S., et al.\ 2022, The Astronomer's Telegram, 15193

\bibitem[Nesci(2022)]{ATel15194.N} Nesci, R.\ 2022, The Astronomer's Telegram, 15194


\bibitem[Pavlinsky et al.(2021)]{2021A&A...650A..42P} Pavlinsky, M., Tkachenko, A., Levin, V., et al.\ 2021, \aap, 650, A42. doi:10.1051/0004-6361/202040265

\bibitem[Pike et al.(2022)]{2022ApJ...927..190P} Pike, S.~N., Negoro, H., Tomsick, J.~A., et al.\ 2022, \apj, 927, 190. doi:10.3847/1538-4357/ac5258

\bibitem[Pradhan et al.(2018)]{2018A&A...610A..50P} Pradhan, P., Bozzo, E., \& Paul, B.\ 2018, \aap, 610, A50. doi:10.1051/0004-6361/201731487


\bibitem[Predehl et al.(2021)]{2021A&A...647A...1P} Predehl, P., Andritschke, R., Arefiev, V., et al.\ 2021, \aap, 647, A1. doi:10.1051/0004-6361/202039313


\bibitem[Reig (2011)]{Reig2011}  
Reig, P.\ 2011, \apss, 332, 1

\bibitem[Rhodes et al.(2022)]{2022ATel15209....1R} Rhodes, L., van den Eijnden, J., Fender, R., et al.\ 2022, The Astronomer's Telegram, 15209

\bibitem[Serino et al.(2022)]{ATel15178.S} 
Serino, M., Negoro, H., Nakajima, M., et al.\ 2022, The Astronomer's Telegram, 15178

\bibitem[Sguera et al.(2006)]{2006ApJ...646..452S} Sguera, V., Bazzano, A., Bird, A.~J., et al.\ 2006, \apj, 646, 452. doi:10.1086/504827

\bibitem[Shakura et al.(2012)]{2012MNRAS.420..216S} Shakura, N., Postnov, K., Kochetkova, A., et al.\ 2012, \mnras, 420, 216. doi:10.1111/j.1365-2966.2011.20026.x

\bibitem[Sidoli \& Paizis(2018)]{2018MNRAS.481.2779S} Sidoli, L. \& Paizis, A.\ 2018, \mnras, 481, 2779. doi:10.1093/mnras/sty2428

\bibitem[Sidoli et al.(2022)]{2022MNRAS.512.2929S} Sidoli, L., Sguera, V., Esposito, P., et al.\ 2022, \mnras, 512, 2929. doi:10.1093/mnras/stac691

\bibitem[Sugizaki et al. (2011)]{Sugizaki2011}
Sugizaki,~M., et al. 2011, \pasj, 63, S635

\bibitem[Sugizaki et al.(2017)]{2017PASJ...69..100S} Sugizaki, M., Mihara, T., Nakajima, M., et al.\ 2017, \pasj, 69, 100.
  
\bibitem[Sunyaev et al.(2021)]{2021A&A...656A.132S} Sunyaev, R., Arefiev, V., Babyshkin, V., et al.\ 2021, \aap, 656, A132. doi:10.1051/0004-6361/202141179

\bibitem[Tsygankov et al.(2021)]{2021arXiv210806365T} Tsygankov, S.~S., Molkov, S.~V., Doroshenko, V., et al.\ 2021, arXiv:2108.06365



\bibitem[Wilms et al.(2000)]{Wilms2000}
Wilms, J., Allen, A., McCray, R.\ 2000, \apj, 542, 914  





\end{thebibliography}
\end{document}